\documentclass[prl,twocolumn,superscriptaddress,nofootinbib,amsmath,amssymb,longbibliography]{revtex4-1}
\usepackage{graphicx}
\usepackage[usenames,dvipsnames]{xcolor}
\usepackage[colorlinks=true,citecolor=Blue,linkcolor=RubineRed,urlcolor=Blue]{hyperref}

\begin{document}

\title{
  Transient Pattern Formation in an Active Matter Contact Poisoning Model
}
\author{ P. Forg{\' a}cs}
\affiliation{Lakeside Labs GmbH, Klagenfurt 9020, Austria}
\affiliation{Physics Department, Babe{\c s}-Bolyai University, Cluj-Napoca 400084, Romania}
\author{A. Lib{\' a}l}
\affiliation{Mathematics and Computer Science Department, Babe{\c s}-Bolyai University, Cluj-Napoca 400084, Romania}
\author{C. Reichhardt}
\affiliation{Theoretical Division and Center for Nonlinear Studies,
Los Alamos National Laboratory, Los Alamos, New Mexico 87545, USA}
\author{N. Hengartner}
\affiliation{Theoretical Division and Center for Nonlinear Studies,
Los Alamos National Laboratory, Los Alamos, New Mexico 87545, USA}
\author{C. J. O. Reichhardt$^*$}
\affiliation{Theoretical Division and Center for Nonlinear Studies,
Los Alamos National Laboratory, Los Alamos, New Mexico 87545, USA}

\date{\today}

\begin{abstract}

One of the most notable features in repulsive particle based active matter systems is motility-induced-phase separation (MIPS) where a dense, often crystalline phase coexists with a low density fluid. In most active matter studies, the activity is kept constant as a function of time; however, there are many examples of active systems in which individual particles transition from living or moving to dead or nonmotile due to lack of fuel, infection, or poisoning. Here we consider an active matter particle system at densities where MIPS does not occur. When we add a small number of infected particles that can effectively poison other particles, rendering them nonmotile, we find a rich variety of time dependent pattern formation, including MIPS, a wetting phase, and a fragmented state formed when mobile particles plow through an nonmotile packing. We map out the patterns as a function of time scaled by the duration of the epidemic, and show that the pattern formation is robust for a wide range of poisoning rates and activity levels. We also show that pattern formation does not occur in a random death model, but requires the promotion of nucleation by contact poisoning. Our results should be relevant to biological and active matter systems where there is some form of poisoning, death, or transition to nonmotility.
\end{abstract}
\maketitle

\section{Introduction}

Active matter denotes systems of interacting particles that benefit from  
some form of self motility.
It has
been used to model
biological systems, active colloids, artificial swimmers, 
social systems, and robotics
\cite{Marchetti13,Bechinger16,Gompper20,Helbing01,Wang21,BenZion23}.
Due to their nonequilibrium nature, active systems
exhibit a variety of irreversible behavior
\cite{Solon15,Takatori14,Dabelow19,OByrne22}, boundary interactions
\cite{Ray14,Ni15,Yan15,Wagner22,Speck20},
and novel transport effects \cite{Reichhardt17a}. 
One of the most striking phenomena in
particle based active matter systems
is motility induced phase separation (MIPS)
\cite{Fily12,Redner13,Palacci13,Buttinoni13,Cates15,Paoluzzi22},
where even for purely steric repulsive interactions that produce
a uniform fluid in the Brownian limit,
phase separation occurs at large activity into a high density solid  
surrounded by a lower density fluid or gas.
The transition to the MIPS phase occurs
as a function of increasing activity and/or particle density.

The propulsion in active systems can arise
from chemical reactions on the surfaces of colloidal particles, 
from biological motors,
or from mechanical motors in robots.
It is generally possible
for individual active
particles to transition into a non-active or 
nonmotile state    
for several reasons, such as exhaustion of available fuel,
finite motor lifetime,
or death of a biological agent due to poisoning or excessive age.
Depending on the scenario, the non-active state may be 
permanent or temporary.
Despite the wide variety of scenarios
in which the lifetime of the activity is finite,
relatively little is known about how this could impact pattern formation.
For example, if the active system is in a regime where MIPS is absent,
reducing the mobility should push the system further away from the MIPS
regime.
If the system is already in a MIPS phase, the addition of dead
or nonmotile particles
may be expected simply to break up the clustering.

Some works have addressed mixtures of active and passive
(nonmotile) particles in
different limits.
For example, a small number of active dopants added to a passive system
can lead to the annealing of topological defects or can cause
the passive particles to segregate into a crystalline-like
arrangement
\cite{Ni14,Kummel15,MassanaCid18,Omar19,Ramananarivo19}.
Other studies of active and passive binary mixtures have
focused on how the active particles
can cause the passive particles to form a dense cluster
\cite{Stenhammar15,Gokhale22}.
Up until now, in studies of this type
the ratio of active to passive particles has been held fixed,
and it is not clear what would happen if there were some
form of poisoning present in the system that could cause
the number of passive particles to increase over time.
Poisoning can be modeled by introducing
nonmotile particles that can spread their
nonmotile state to mobile particles through
direct contact to create an epidemic-like transmission.
A variant of this is to introduce a
random death process where individual
particles transition suddenly from mobile to
nonmotile irrespective of any interactions with other particles.
In both of these
cases, mobile particles are only transiently present, and in the final
state all of the particles are nonmotile.
An open question is whether any kind of pattern formation
occurs in such systems and whether it makes a difference if 
nonmotility
is introduced through contact poisoning or through random death.
   
Motivated by recent
models of epidemic behavior in active  matter systems for
susceptible-infected-recovered (SIR)
dynamics \cite{Paoluzzi20,Zhao22,Forgacs22,Libal23},
in this work we study active run-and-tumble disks
under non-MIPS conditions where we
add a small number of nonmotile
or dead particles that act as poisoning agents.
With some finite probability, a mobile particle that is in contact
with a nonmotile particle becomes nonmotile.
We find that despite the apparent
simplicity of this model, it exhibits a wide variety  of
transient pattern formation.
At early times, the dead particles create small local clusters
that act as nucleation sites
for the growth of a large dense triangular cluster consisting of
a dead central region surrounded by an accumulation of
motile active particles.
The accumulation process can be viewed
as an example of active wetting or aggregation
along walls, as observed
in previous studies of active matter systems with barriers
\cite{Tailleur09,Sepulveda17,Neta21,Turci21}.
The active particles in the accumulation region are progressively
converted to nonmotile particles, driving the growth of the
dense crystalline cluster.
Once the nonmotile dense cluster has consumed a sufficiently large fraction
of particles, 
the remaining active particles begin to
create a fragmented zone along the edges of the cluster.
The fragments consist of an intermediate density
amorphous packing of passive particles crisscrossed by
plow tracks from the active particles.
The fragmented state penetrates
the dense phase as a well defined front
until the dense phase has become entirely fragmented.

Signatures of the pattern formation appear as
a nonmonotonic time dependent behavior
of the cluster size and the
average number of contacts.
We map out the transient cluster formation as a function
of the activity level or run length and the 
poisoning probability $\beta$,
and show that the pattern formation
we observe is robust for a wide range of parameters.
For low activity or low infectivity,
pattern formation is absent.
In a higher density system that can form MIPS states,
the poisoning breaks up the MIPS.
We show that the pattern formation we observe is
due to the particle-particle information
exchange, where the initial clustering of dead particles
produces nucleation sites
that cause active particles
to become nonmotile or effectively stick to the cluster.
In systems with random death, the pattern formation is lost.
Our research reveals  qualitative properties
that can be hypothesized and tested in
active biological systems
such as bacteria
that die after infection by bacteriophages
and act as a poison to surrounding
bacteria \cite{Styles21}.
Other experiments could be performed using
active colloids with optical feedback effects
\cite{Pince16,Lavergne19,Bauerle20}.

\section{Methods}
In previous work,
we studied epidemic spreading in active matter systems
with SIR models
\cite{Kermack27,Hethcote00} or variations thereon \cite{Forgacs22,Libal23}.
We considered a case where a susceptible (S) active particle comes into
contact with an infected (I) particle, and during each simulation time
step of contact,
there is a probability $\beta$ that the S particle will convert to I.
If the only available states are susceptible and infected, the model
is known as the susceptible-infected (SI) model,
and,
provided that every particle can come in contact with other particles,
all particles will eventually transform to I.
In the SIR model, I particles spontaneously
transition to the recovered (R) state with a rate $\mu$,
opening the possibility that all the infected particle recover before the pool
of susceptible particles is exhausted.
As a result, over a range of $\beta$ and $\mu$ values,
a portion of the initial
S particles never become infected.
The epidemic size,
defined as the fraction of recovered particles at the end of the epidemic,
depends on 
the ratio of the transmission and recovery rates.
In previous studies of SI and SIR epidemic 
dynamics on active matter,
the infected particles
remained mobile;
however, in many
physically relevant cases, infected particles would likely become nonmotile.
In biological systems,
a more realistic
situation could be that
the infected particles die, but remain capable of
transmitting the infection or acting as a
poison to other particles.

Here we consider a variation of the active matter SI model
that we term the
contact poisoning model.
In this model, S particles are capable of becoming infected (poisoned),
I denotes a nonmotile infectious or
poisoning particle, and there is no spontaneous recovery.
In this model, all the particles are eventually infected.
We consider a two-dimensional system
with periodic boundary conditions in
the $x$ and $y$ directions
of size $L \times L$ with $L=200.$
We typically focus on
samples containing $N=4000$ run-and-tumble active particles,
which are modeled as
harmonically repulsive disks of radius
$r_a=1.0$.
Unless otherwise noted,
the disk density is $\phi =  N \pi r_a^2/L^2 = 0.31415$,
low enough that
the system
does not enter the MIPS phase for the activity levels we consider.

The particles obey overdamped dynamics
in which the time evolution
of disk $i$ is governed by the following equation of motion:
\begin{equation} 
\alpha_d {\bf v}_{i}  =
{\bf F}^{dd}_{i} + {\bf F}^{m}_{i} \ .
\end{equation}
Here the particle velocity is  ${\bf v}_{i} = {d {\bf r}_{i}}/{dt}$,
${\bf r}_{i}$ is the location of particle $i$,
and the damping constant $\alpha_d = 1.0$.
The disks interact sterically via a repulsive harmonic potential
${\bf F}^{dd}_{i} = \sum_{i\neq j}^{N}k(|{\bf r}_{ij}| - 2r_a)\Theta( |{\bf r}_{ij}| - 2r_{a}) {\hat {\bf r}_{ij}}$, where $\Theta$ is the Heaviside step function, ${\bf r}_{ij} = {\bf r}_{i} - {\bf r}_{j}$, $\hat {\bf r}_{ij}  = {\bf r}_{ij}/|{\bf r}_{ij}|$, and the repulsive spring force constant is
$k = 20.$
The activity arises from a motor
force ${\bf F}_i^m=F_{M}{\bf \hat{m}}_i$ of magnitude $F_M$
applied in a randomly chosen direction ${\bf \hat{m}}_i$.
A new random direction of motion is selected after each time period
$\tau_l \in [\tau, 2\tau]$.
This model has been used previously to identify the transition
to a MIPS phase and to study active matter versions of SIR.
\cite{Forgacs22,Libal23,Reichhardt14,Sandor17a}.

Using the results from our previous work, we select
a run length and motor force for which the system is in a uniform
density phase rather than
in the MIPS regime.
On top of the discrete dynamics described above, each particle
carries a variable indicating whether it is currently in the S or I state.
Particles in the I state have $F_M$ set to $F_M=0$, rendering their
motor inoperative; however, these particles can still move if pushed
by direct contact with active particles.
After each simulation time step update of the particle positions using
Eq.~(1), we update the status of the mobile active particles. 
For each pair of  an I particle in direct contact 
with an S particle (i.e., $| {\bf r}_{ij}| \leq 2r_a$), 
the S particle will change to I during that simulation
time step with probability $\beta$. 
This adds an information exchange dynamic
onto the active matter system.
We consider values of $\beta$ spanning the range
$2\times 10^{-6} \leq \beta \leq 3.2\times 10^{-5}$.
To initialize the system, we place all particles at randomly chosen
nonoverlapping positions and set all of the particles to S except for
five randomly chosen poisoned I particles.

\section{Results}

\begin{figure}
\includegraphics[width=0.5\textwidth]{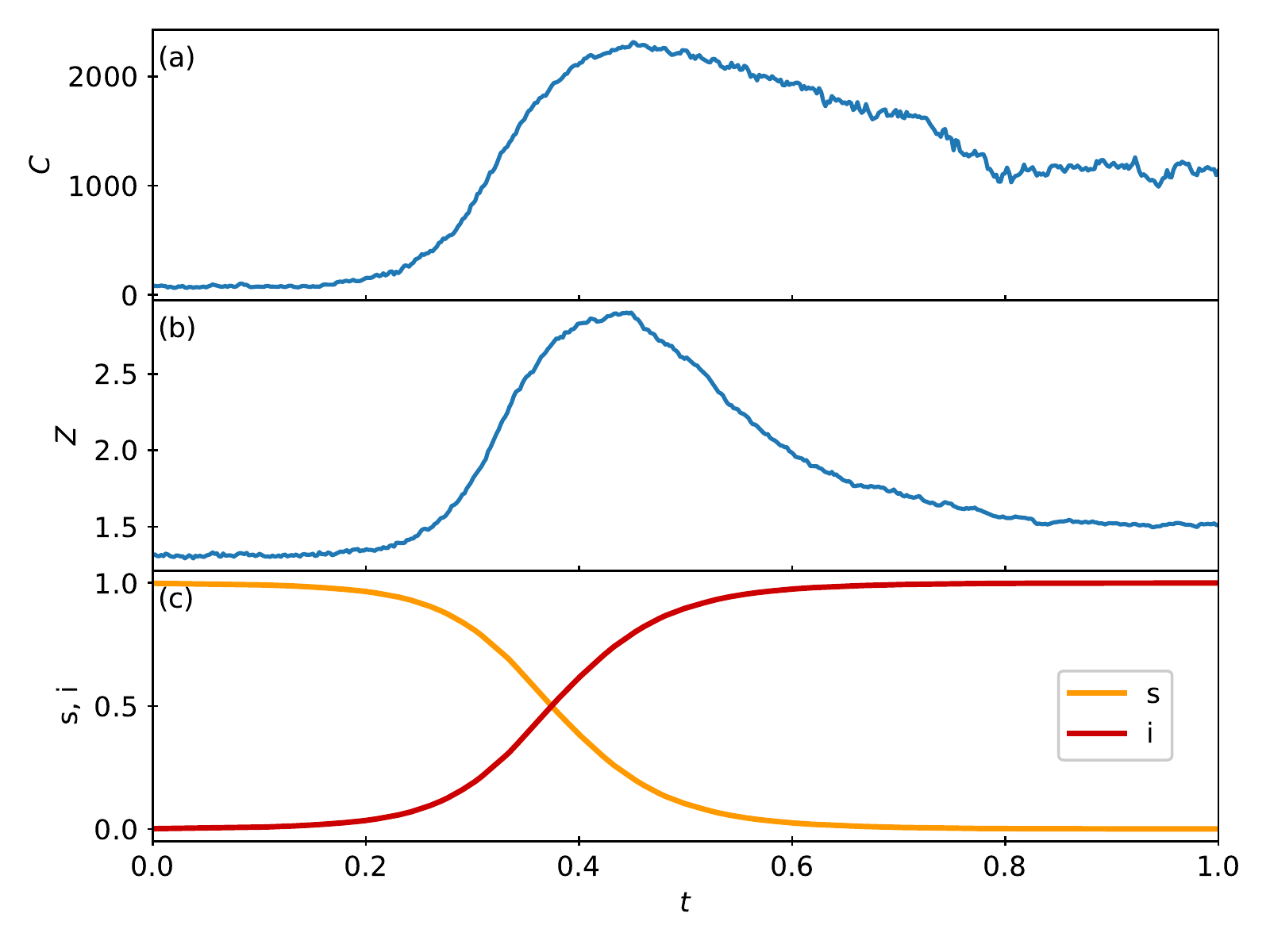}
\caption{ Time evolution of
(a) cluster size $C$ or number of particles in the largest cluster,
(b) the average contact number $Z$ per particle,
and (c) the fraction $s$ of susceptible particles (yellow)
and infected nonmotile particles $i$ (red)
for a system with
$\phi =  0.31415$,
$\tau=15000$,
$F_{M} = 1.5$,
and $\beta = 8 \times 10 ^{-6}$.
The curves are averaged over ten realizations.
}
\label{fig:1}
\end{figure}

We summarize the dynamics of the epidemic using the temporal
evolution of the fraction
of infected individuals $i(t)$. We want to relate this
to the time evolution of the largest cluster size $C$
and the average contact number $Z$ per particle,
both of which are determined by
examining all particles in direct contact with each other.
In Fig.~\ref{fig:1}(a,b),
we show an example of how $C$ and $Z$ evolve
for a system with
$\phi =  0.31415$,
$\tau=15000$,
$F_{M} = 1.5$,
and $\beta = 8 \times 10 ^{-6}$.
The time $t$ is normalized by the duration of the epidemic, defined as the
time required for the last S particle to reach state I.
Figure~\ref{fig:1}(c) shows
the corresponding fraction $s$ of susceptible particles
and infected nonmotile particles $i$.
The data is obtained by averaging together ten different realizations
in which time has been normalized by the duration of the epidemic, defined
as the time required for the last S particle to reach state I so that
all particles are nonmotile.
Here, $C$ and $Z$ are nonmonotonic
and reach a peak value near $t = 0.45$, indicative of
transient cluster formation.

\begin{figure}
\includegraphics[width=0.5\textwidth]{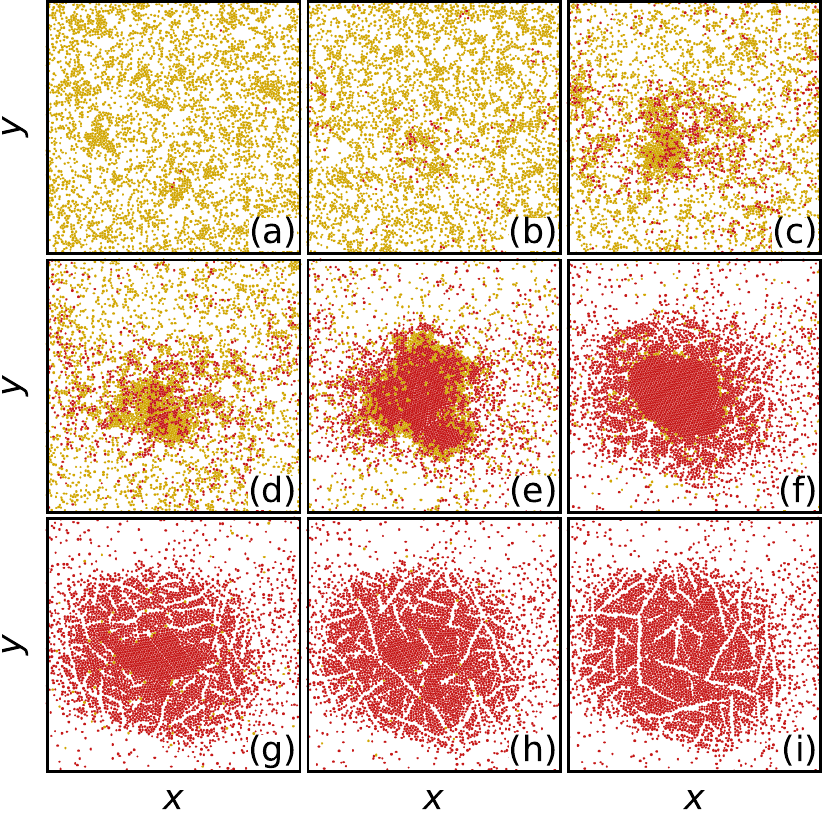}
\caption{ The particle locations (yellow: S or mobile; red: I or infected
nonmotile)
at different times $t$
for the system
in Fig.~\ref{fig:1} with
$\phi =  0.31415$,
$\tau=15000$,
$F_{M} = 1.5$,
and $\beta = 8 \times 10 ^{-6}$.
(a) $t = 0.017$.
(b) $t = 0.214$.
(c) $t = 0.325$.
(d) $t = 0.342$.
(e) $t = 0.410$.
(f) $t = 0.504$.
(g) $t = 0.590$.
(h) $t = 0.650$.
(i) $t = 1.0$ showing the final fragmented state.
At early times (a,b,c,d), a dense crystalline cluster nucleates and grows,
while at later times (e,f,g), fragmentation of the dense cluster occurs.
For very late times in (h,i),
a plowing effect appears.
A video of this process appears in the supplemental material \cite{Suppl}.
}
\label{fig:2}
\end{figure}

In Fig.~2 we illustrate the positions of the S and I particles at different
times for the system from Fig.~\ref{fig:1}.
At $t = 0.017$ in Fig.~\ref{fig:2}(a),
most of the particles are still in state S and form
a weakly clustered fluid
in which MIPS does not occur.
In Fig.~\ref{fig:2}(b) at $t = 0.214$,
larger clusters are beginning to nucleate around
the growing number of I particles,
but the clusters are amorphous and fluctuate with time.
At $t = 0.325$ in Fig.~\ref{fig:2}(c) and $t=0.342$ in Fig.~\ref{fig:2}(d)
a dominant crystalline cluster containing a mixture of S and I particles
is emerging,
while at $t = 0.410$ in Fig.~\ref{fig:2}(e),
there is now a dense triangular
lattice composed entirely of I at the center of the cluster
surrounded by a wetting layer of S on the outer edge of the cluster.
In Fig.~\ref{fig:2}(f) at $t = 0.504$,
the system consists of a dense triangular core of nonmotile particles
with a halo-like region of lower density amorphous phase
containing tracks, which we call a fragmented
phase.
This fragmented phase
moves into the dense region as a well-defined front
driven by the plowing effect of the remaining mobile S particles.
At
$t = 0.590$ in Fig.~\ref{fig:2}(g),
the dense core has been almost completely depleted and
replaced by the fragmented phase.
For $t = 0.65$ in Fig.~\ref{fig:2}(h),
the system is completely fragmented and contains only a small remnant
of moving S particles, while Fig.~\ref{fig:2}(i) shows the $t=1.0$ final
state
in which all of the particles are dead
and form low density amorphous clusters that are separated by tracks.

From the time-dependent images and behavior of $C$ and $Z$,
we can identify several different regimes.
For  $0 < t < 0.25$  where $Z$, $C$ and $i$ 
are all low,
only small clusters are present, as shown in
Fig.~\ref{fig:2}(a,b),
while for $0.25 \leq t < 0.45$,
we see the emergence of a large dense cluster
reflected by  the increase
of $C$ and $Z$ in Fig.~\ref{fig:1}(a,b).
The cluster reaches its maximum size near
the point at which $i=2/3$ and $s=1/3$.
For $0.45 \leq t < 0.75$, the fragmentation region
begins to form around the dense
zone, resulting in a drop in $C$ and $Z$ in
Fig.~\ref{fig:1}(a,b),
while for $t \geq 0.75$,
the system is completely fragmented and contains only
a small number of mobile S
particles.

\begin{figure}
\includegraphics[width=0.5\textwidth]{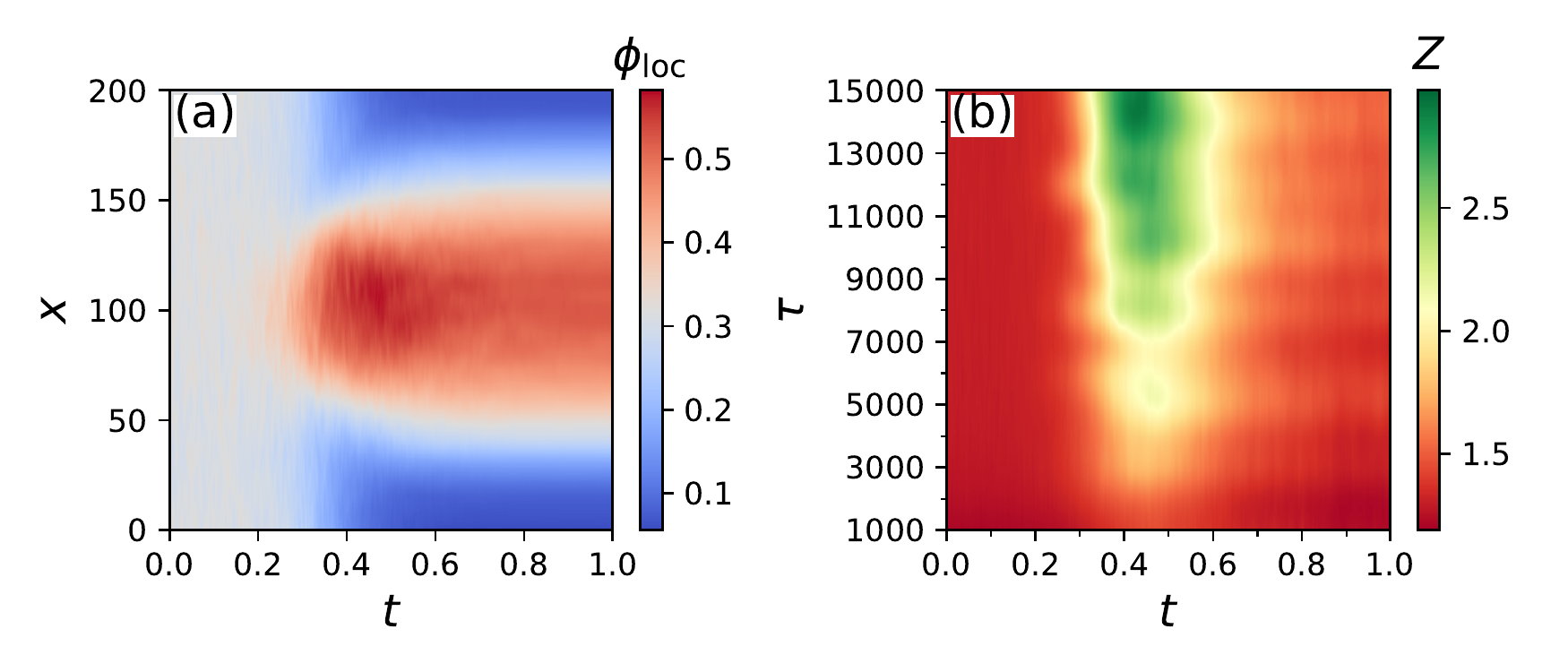}
\caption{
(a) Heat map of the local density $\phi_{\rm loc}$ at a fixed
value of $y$ near the center of the system plotted as a function
of $x$ position versus time $t$
for the system in Fig.~\ref{fig:2} with
$\phi =  0.31415$,
$\tau=15000$,
$F_{M} = 1.5$,
and $\beta = 8 \times 10 ^{-6}$,
showing the formation of the cluster.
(b) Phase diagram in the form of a heat map of $Z$ plotted as a
function of $\tau$ versus $t$
for the system in Fig.~\ref{fig:1}
with $\phi = 0.31415$ and $F_M=1.5$.
Clustering is present for sufficiently large $\tau$
only at intermediate times, as indicated by the green region.
}
\label{fig:3}
\end{figure}

To further illustrate the behavior, we measure the
time evolution of the local particle density $\phi_{\rm loc}$ in a
slice of the sample taken along the $y$ direction.
We plot a height map of $\phi_{\rm loc}$ as a function of
$x$ position versus time $t$ in
Fig.~\ref{fig:3}(a) for the system
from Fig.~\ref{fig:2}.
The density remains uniform up to $t = 0.2$,
but at later times
a strong density peak appears in the center
of the sample, corresponding to the
formation of the dense crystalline phase.
At larger times $t$, the fragmented state 
becomes 
visible 
with the appearance of strips of low density along the edges of the dense region.

By conducting
a series of simulations 
at varied
$\tau$, we construct the phase
diagram shown in Fig.~\ref{fig:3}(b)
in the form of a height field of the average coordination number $Z$
plotted as a function of $\tau$ versus $t$.
For $\tau > 5000$, the system can form a dense
phase, indicated by the green region,
while for $\tau \leq  5000$ the system remains in a uniform
phase.
This indicates that Brownian particles obeying the same S to I dynamics
will only form a uniform nonmotile state,
which we interpret to mean
that the activity is essential for producing pattern formation.

\begin{figure}
\includegraphics[width=0.5\textwidth]{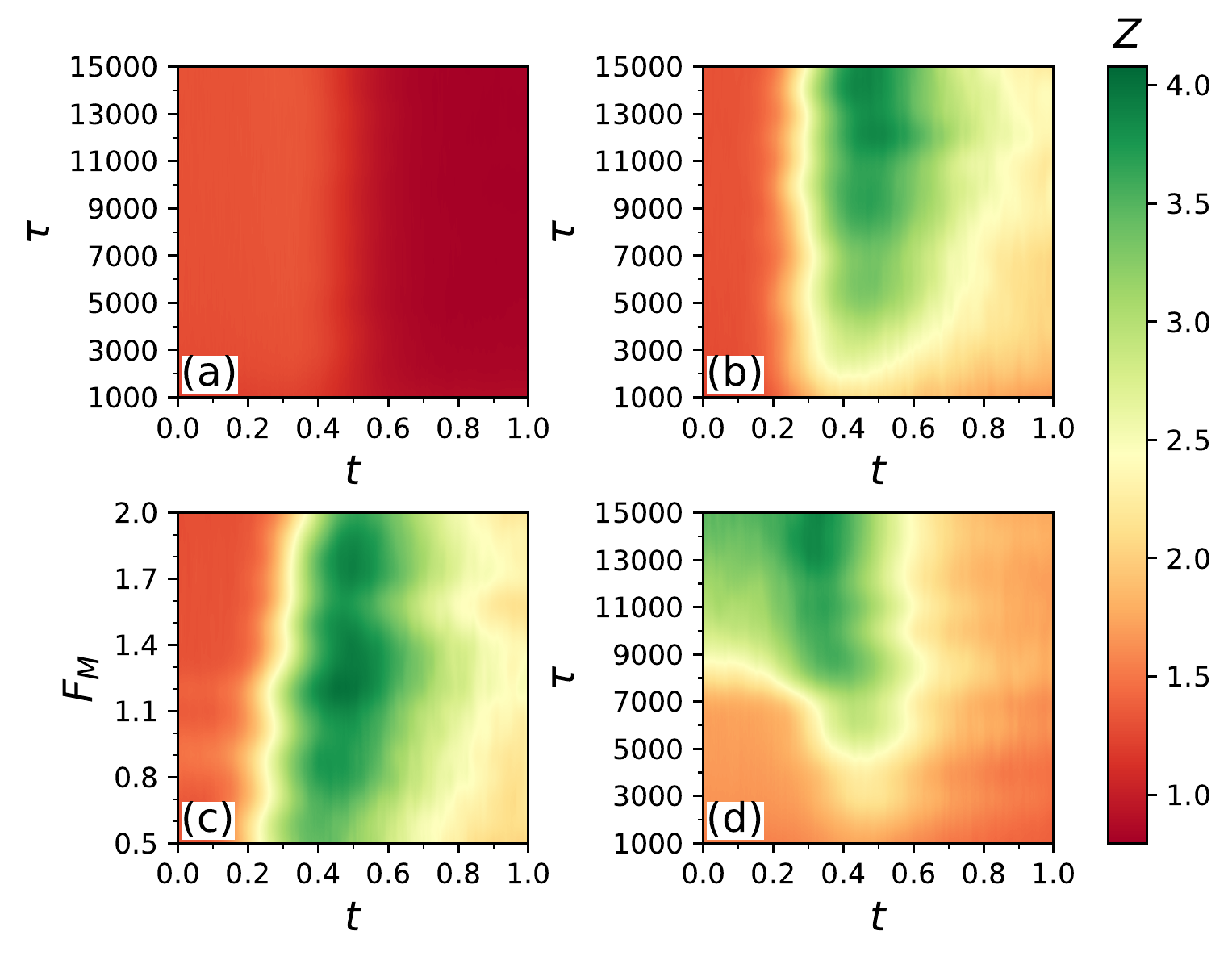}
\caption{
Phase diagrams in the form of heat maps of $Z$.  
(a) $Z$ as a function of $\tau$ vs $t$ for
the same system in Fig.~\ref{fig:3}(b)
with $\phi=0.31415$
and $F_M=1.5$
at
a lower infection rate of
$\beta = 2\times 10^{-6}$.
where the amount of clustering
is reduced. 
(b) The same for
a system with a higher infection rate $\beta = 3.2\times 10^{-5}$, where
the clustering is enhanced.
(c) $Z$ as a function of
motor force $F_{M}$ vs $t$
for a system with
$\phi = 0.31415$
and
$\tau = 15000$
at
$\beta = 3.2\times 10^{-5}$.	
(d) $Z$ as a function of $\tau$ vs $t$ for a higher $\phi = 0.393$ at
$F_M=1.5$ and 
$\beta = 8.0\times 10^{-6}$. At this density, MIPS clustering occurs
at $t=0$
for larger values of $\tau$, but the poisoning process still destroys
the cluster over time.}
\label{fig:4}
\end{figure}

We tested the impact of the
poisoning probability
on the pattern formation,
as shown in Fig.~\ref{fig:4}(a)
where we plot a height field of $Z$ as a function of $\tau$ versus $t$ for
the same system as in Fig.~\ref{fig:3}(b) but for
a lower $\beta = 2\times 10^{-6}$.
Here the clustering is strongly reduced. If we instead 
increase $\beta$ to 
$\beta = 3.2\times 10^{-5}$ for the same system, the clustering is
strongly enhanced, as shown in
the heat map of $Z$ in Fig.~\ref{fig:4}(b).
In general, if  $\beta$ is very  large,
it is possible for a portion of the
dense phase to persist to the end of the epidemic
since the mobile particles become infected too quickly to be
able to
fragment the system.

Figure~\ref{fig:4}(c) shows the $Z$ heat map phase diagram as a 
function of
$F_{M}$ versus $t$
for a system with
$\phi = 0.31415$
and
$\tau = 15000$
at
$\beta = 3.2\times 10^{-5}$,
where a cluster always forms at intermediate time and the fragmented
phase appears at late time.
We have also considered a system with higher $\phi$ where a MIPS
state appears at $t=0$ for sufficiently large $\tau$
even without poisoning.
We first allow the system to organize into a MIPS state
prior to adding the poisoned particles.
When the $t=0$ MIPS state is present,
the addition of poisoning
dynamics only slightly increases the amount of clustering
at intermediate times, but
in general, the poisoning breaks up the MIPS at later times.
In Fig.~\ref{fig:4}(d)
we show the $Z$ heat map phase diagram
as a function of $\tau$ versus $t$
for a higher $\phi = 0.393$ at
$\beta = 8.0\times 10^{-6}$.
Here, for $\tau > 9000$ the system starts in a MIPS phase
and shows a weak enhancement of the clustering prior to the
onset of fragmentation,
while for $4000 < \tau < 9000$ 
the system transitions from a fluid cluster to a fragmented state.
For $\tau < 4000$, the system is always in a fluid phase.

\begin{figure}
\includegraphics[width=0.5\textwidth]{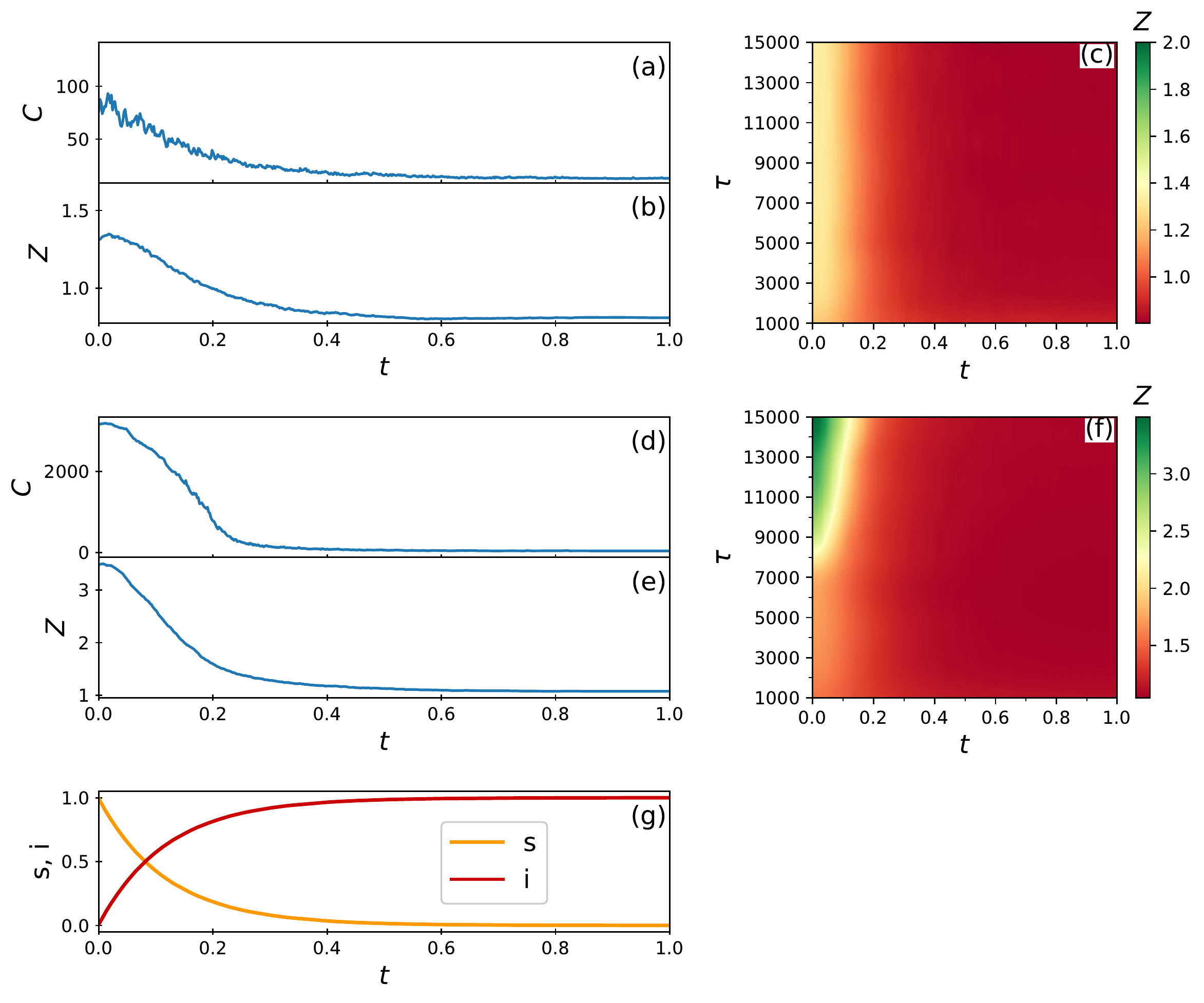}
\caption{ 
Results from the sudden death model,
where S particles transition to I
with probability
$\alpha = 8\times 10^{-6}$ per time step, independent of particle-particle
interactions.
(a) $C$ and (b) $Z$
vs $t$ for a sample with  
$\phi = 0.31415$
and
$\tau = 15000$.
(c) The corresponding phase diagram in the form of a height map of $Z$ as
a function of $\tau$ vs $t$,
showing that there is no pattern formation.
(d) $C$ and (e) $Z$
vs $t$ for a sample with  
$\phi = 0.393$ and $\tau = 15000$, where a MIPS state forms
and the random deaths simply reduce the clustering.
(f) The corresponding phase diagram in the form of a height map of $Z$
indicates that the random death model breaks down
existing clustering and produces no new clustering.
(g) The $s$ (yellow) and $i$ (red) curves vs $t$
are the same for both
sudden death samples.
}
\label{fig:5}
\end{figure}

We have also considered the case of sudden random death,
where there is no interaction between S and I particles, but instead S
particles spontaneously transition to I with probability $\alpha$.
In Fig.~\ref{fig:5}(a,b) we plot
$C$ and $Z$ versus time for a sudden death
sample with $\phi = 0.31415$
and $\tau = 15000$,
which corresponds to a regime where transient clustering occurs for
the poisoning model.
The probability for any single particle to die in a
given time step is $\alpha = 8\times 10^{-6}.$
The $s$ and $i$ vs $t$ curves appear in Fig.~\ref{fig:5}(g).
On average, the amount of simulation
time required to transform the last S particle into I
is much larger than for the contact poisoning
model, leading to a more asymmetric shape of the curves as a function
of rescaled time $t$.
There are two effects contributing to this.  First, in an SI model,
particles making the greatest number of contacts are more likely to come
into contact with infected particles and become infected more quickly, 
while the last surviving particles are typically the ones that
have made the least contact with the infected particles.
Such contact dynamics are irrelevant in the random 
death model.
Second,
there is no formation of a dense triangular phase
in the random death model,
and the largest cluster that forms is only of size $C=80$.
In Fig.~\ref{fig:5}(c), we plot the sudden death phase
diagram in the form of a height field of $Z$
as a function of $\tau$ versus $t$
for the same system as in Fig.~\ref{fig:5}(a,b),
indicating the lack of pattern formation over this range of $\tau$.
In Fig.~\ref{fig:5}(d,e) we plot $C$ and $Z$ versus $t$ for
a sudden death sample with the same $\tau=15000$ but at a higher
density of $\phi = 0.393$.
The $s$ and $i$ versus $t$ curves are unaffected by $\phi$ and appear
in Fig.~\ref{fig:5}(g). In this case,
the system starts in a MIPS phase at $t = 0$, and the sudden
death breaks up the cluster without creating any new clustering.
The height field plot of $Z$ as a function of $\tau$ versus $t$ in
Fig.~\ref{fig:5}(f) for the $\phi=0.393$ sample indicates
that the transient clustering found under interacting SI dynamics
does not occur in the sudden death model.

\begin{figure}
\includegraphics[width=0.5\textwidth]{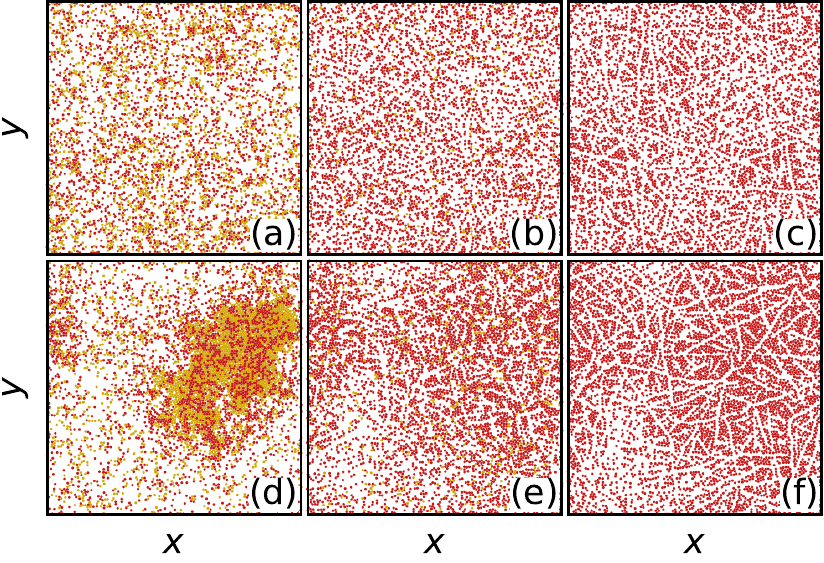}
\caption{The particle positions (yellow: S or mobile; red: I or infected
nonmotile)
for the sudden death system from Fig.~\ref{fig:5} with $\tau=15000$.
(a,b,c) The lower density system with $\phi=0.31415$
from Fig.~\ref{fig:5}(a-c)
at (a) $ t = 0.1$, (b) $t = 0.3$, and (c) $t =1.0$. 
(d,e,f) The higher density system with $\phi=0.393$ from Fig.~\ref{fig:5}(d-f)
at (d) $t = 0.1$, (e) $t = 0.3$, and (f) $t = 1.0$. 
Movies of the two cases are available in the Supplemental Material \cite{Suppl}.
}
\label{fig:6}
\end{figure}

Figure~\ref{fig:6}(a,b,c) illustrates the particle configurations
in the sudden death model for the
lower density $\phi=0.31415$
system in
Fig.~\ref{fig:5}(a-c).
At $t = 0.1$ in Fig.~\ref{fig:6}(a), small clusters appear
throughout the sample.
In Fig.~\ref{fig:6}(b) at $t=0.3$,
the clusters have decreased in size and the number of I particles has
increased, while
Fig.~\ref{fig:6}(c) shows the final state
in which I particles are spread everywhere throughout the sample separated
by voids marking the plow tracks of the final S particles.
The $\phi=0.393$ sample from
Fig.~\ref{fig:5}(d,e,f) is illustrated in Fig.~\ref{fig:6}(d,e,f).
At $t=0.1$  in Fig.~\ref{fig:6}(c), the system has not yet
changed much from its initial
MIPS state, but the MIPS cluster contains a moderate density of infected
nonmotile
I particles.
In Fig.~\ref{fig:6}(e) at $t= 0.3$,
the MIPS cluster is disintegrating,
and in the final $ t = 1.0$ configuration in Fig.~\ref{fig:6}(f),
a lower density fragmented state appears.
These results indicate that contact poisoning is essential for producing
the transient clustering, and that the random death model does not give
the same types of patterns.
This is also consistent with the loss of patterning
in the contact poisoning model for low $\beta$,
shown in Fig.~\ref{fig:4}(a), where the odds of
infection become low so that
even if an S particle encounters a group of poisoned I particles,
the S particle has a high chance of moving
away before it becomes poisoned.
For high values of $\beta$, shown in Fig.~\ref{fig:4}(b),
the S particle has a higher chance to become
infected and effectively stick to a cluster of dead I particles,
which serve as cluster nucleation sites.
In this way, the low infectivity
limit is closer to the random death model.

\section{Discussion}

In our model, even though the system is not
in a MIPS regime, it can form a transient cluster similar to a MIPS
phase that becomes fragmented by the remaining active particles.
There are some differences between MIPS and the transient clustering.
The active particles do not produce the transient clustering;
instead, it is the result of the poisoning by dead I particles
that act as nucleation sites and form dense regions in which
the odds of a mobile S particle becoming poisoned are high.
In contrast, for random death models
where interactions between particles play no role in the transition
to the I state, we do not observe any transient clustering.
Our contact poisoning model
should be relevant to physical active systems, since most such systems
contain some constraint that can stop the mobility. This could
include local depletion of resources needed for motion.
A scenario in which individual particles carry their own source of fuel
that can become exhausted would be more
consistent with the random death model than with contact poisoning.

Another variation to consider is activation rather than poisoning,
in which all of the particles are initially nonmotile but introduction
of a small number of active particles can lead to the activation of
additional particles through contact interactions.
A competing scenario would be one in which nonmotile particles become
active with some probability per time step, independent of interactions
with other particles; this would be a sudden life model.
In both cases, the final state would be that all particles would reach
state S.
It would be interesting
to see whether the same sets of phases occur
in reverse order or what
would be the effect of adding spatial inhomogeneities 
\cite{Morin17,Bhattacharjee19a,Forgacs21}.
Finally, an active matter SIS model in which infected particles stop while infected and then restart moving
when recovered presents a third scenario that combines both of the previous ones.  We surmise that the
latter may, once suitably rescaled, behave as a more standard quenched active matter system.

\smallskip

\begin{acknowledgments}
This work was supported by the US Department of Energy through
the Los Alamos National Laboratory.  Los Alamos National Laboratory is
operated by Triad National Security, LLC, for the National Nuclear Security
Administration of the U. S. Department of Energy (Contract No. 892333218NCA000001).
NH benefited from resources provided by the Center for Nonlinear Studies (CNLS).
AL was supported by a grant of the Romanian Ministry of Education
and Research, CNCS - UEFISCDI, project number
PN-III-P4-ID-PCE-2020-1301, within PNCDI III.
\end{acknowledgments}

\bibliography{mybib}

\begin{thebibliography}{49}%
\makeatletter
\providecommand \@ifxundefined [1]{%
 \@ifx{#1\undefined}
}%
\providecommand \@ifnum [1]{%
 \ifnum #1\expandafter \@firstoftwo
 \else \expandafter \@secondoftwo
 \fi
}%
\providecommand \@ifx [1]{%
 \ifx #1\expandafter \@firstoftwo
 \else \expandafter \@secondoftwo
 \fi
}%
\providecommand \natexlab [1]{#1}%
\providecommand \enquote  [1]{``#1''}%
\providecommand \bibnamefont  [1]{#1}%
\providecommand \bibfnamefont [1]{#1}%
\providecommand \citenamefont [1]{#1}%
\providecommand \href@noop [0]{\@secondoftwo}%
\providecommand \href [0]{\begingroup \@sanitize@url \@href}%
\providecommand \@href[1]{\@@startlink{#1}\@@href}%
\providecommand \@@href[1]{\endgroup#1\@@endlink}%
\providecommand \@sanitize@url [0]{\catcode `\\12\catcode `\$12\catcode
  `\&12\catcode `\#12\catcode `\^12\catcode `\_12\catcode `\%12\relax}%
\providecommand \@@startlink[1]{}%
\providecommand \@@endlink[0]{}%
\providecommand \url  [0]{\begingroup\@sanitize@url \@url }%
\providecommand \@url [1]{\endgroup\@href {#1}{\urlprefix }}%
\providecommand \urlprefix  [0]{URL }%
\providecommand \Eprint [0]{\href }%
\providecommand \doibase [0]{http://dx.doi.org/}%
\providecommand \selectlanguage [0]{\@gobble}%
\providecommand \bibinfo  [0]{\@secondoftwo}%
\providecommand \bibfield  [0]{\@secondoftwo}%
\providecommand \translation [1]{[#1]}%
\providecommand \BibitemOpen [0]{}%
\providecommand \bibitemStop [0]{}%
\providecommand \bibitemNoStop [0]{.\EOS\space}%
\providecommand \EOS [0]{\spacefactor3000\relax}%
\providecommand \BibitemShut  [1]{\csname bibitem#1\endcsname}%
\let\auto@bib@innerbib\@empty
\bibitem [{\citenamefont {Marchetti}\ \emph {et~al.}(2013)\citenamefont
  {Marchetti}, \citenamefont {Joanny}, \citenamefont {Ramaswamy}, \citenamefont
  {Liverpool}, \citenamefont {Prost}, \citenamefont {Rao},\ and\ \citenamefont
  {Simha}}]{Marchetti13}%
  \BibitemOpen
  \bibfield  {author} {\bibinfo {author} {\bibfnamefont {M.~C.}\ \bibnamefont
  {Marchetti}}, \bibinfo {author} {\bibfnamefont {J.~F.}\ \bibnamefont
  {Joanny}}, \bibinfo {author} {\bibfnamefont {S.}~\bibnamefont {Ramaswamy}},
  \bibinfo {author} {\bibfnamefont {T.~B.}\ \bibnamefont {Liverpool}}, \bibinfo
  {author} {\bibfnamefont {J.}~\bibnamefont {Prost}}, \bibinfo {author}
  {\bibfnamefont {M.}~\bibnamefont {Rao}}, \ and\ \bibinfo {author}
  {\bibfnamefont {R.~A.}\ \bibnamefont {Simha}},\ }\bibfield  {title} {\enquote
  {\bibinfo {title} {Hydrodynamics of soft active matter},}\ }\href {\doibase
  10.1103/RevModPhys.85.1143} {\bibfield  {journal} {\bibinfo  {journal} {Rev.
  Mod. Phys.}\ }\textbf {\bibinfo {volume} {85}},\ \bibinfo {pages}
  {1143--1189} (\bibinfo {year} {2013})}\BibitemShut {NoStop}%
\bibitem [{\citenamefont {Bechinger}\ \emph {et~al.}(2016)\citenamefont
  {Bechinger}, \citenamefont {Di~Leonardo}, \citenamefont {L\"owen},
  \citenamefont {Reichhardt}, \citenamefont {Volpe},\ and\ \citenamefont
  {Volpe}}]{Bechinger16}%
  \BibitemOpen
  \bibfield  {author} {\bibinfo {author} {\bibfnamefont {C.}~\bibnamefont
  {Bechinger}}, \bibinfo {author} {\bibfnamefont {R.}~\bibnamefont
  {Di~Leonardo}}, \bibinfo {author} {\bibfnamefont {H.}~\bibnamefont
  {L\"owen}}, \bibinfo {author} {\bibfnamefont {C.}~\bibnamefont {Reichhardt}},
  \bibinfo {author} {\bibfnamefont {G.}~\bibnamefont {Volpe}}, \ and\ \bibinfo
  {author} {\bibfnamefont {G.}~\bibnamefont {Volpe}},\ }\bibfield  {title}
  {\enquote {\bibinfo {title} {Active particles in complex and crowded
  environments},}\ }\href {\doibase 10.1103/RevModPhys.88.045006} {\bibfield
  {journal} {\bibinfo  {journal} {Rev. Mod. Phys.}\ }\textbf {\bibinfo {volume}
  {88}},\ \bibinfo {pages} {045006} (\bibinfo {year} {2016})}\BibitemShut
  {NoStop}%
\bibitem [{\citenamefont {Gompper}\ \emph {et~al.}(2020)\citenamefont
  {Gompper}, \citenamefont {Winkler}, \citenamefont {Speck}, \citenamefont
  {Solon}, \citenamefont {Nardini}, \citenamefont {Peruani}, \citenamefont
  {L{\" o}wen}, \citenamefont {Golestanian}, \citenamefont {Benjamin~Kaupp},
  \citenamefont {Alvarez}, \citenamefont {Ki{\o}rboe}, \citenamefont {Lauga},
  \citenamefont {Poon}, \citenamefont {DeSimone}, \citenamefont {Mui{\~
  n}os-Landin}, \citenamefont {Fischer}, \citenamefont {S{\" o}ker},
  \citenamefont {Cichos}, \citenamefont {Kapral}, \citenamefont {Gaspard},
  \citenamefont {Ripoll}, \citenamefont {Sagues}, \citenamefont
  {Doostmohammadi}, \citenamefont {Yeomans}, \citenamefont {Aranson},
  \citenamefont {Bechinger}, \citenamefont {Stark}, \citenamefont {Hemelrijk},
  \citenamefont {Nedelec}, \citenamefont {Sarkar}, \citenamefont {Aryaksama},
  \citenamefont {Lacroix}, \citenamefont {Duclos}, \citenamefont {Yashunsky},
  \citenamefont {Silberzan}, \citenamefont {Arroyo},\ and\ \citenamefont
  {Kale}}]{Gompper20}%
  \BibitemOpen
  \bibfield  {author} {\bibinfo {author} {\bibfnamefont {G.}~\bibnamefont
  {Gompper}}, \bibinfo {author} {\bibfnamefont {R.~G.}\ \bibnamefont
  {Winkler}}, \bibinfo {author} {\bibfnamefont {T.}~\bibnamefont {Speck}},
  \bibinfo {author} {\bibfnamefont {A.}~\bibnamefont {Solon}}, \bibinfo
  {author} {\bibfnamefont {C.}~\bibnamefont {Nardini}}, \bibinfo {author}
  {\bibfnamefont {F.}~\bibnamefont {Peruani}}, \bibinfo {author} {\bibfnamefont
  {H.}~\bibnamefont {L{\" o}wen}}, \bibinfo {author} {\bibfnamefont
  {R.}~\bibnamefont {Golestanian}}, \bibinfo {author} {\bibfnamefont
  {U.}~\bibnamefont {Benjamin~Kaupp}}, \bibinfo {author} {\bibfnamefont
  {L.}~\bibnamefont {Alvarez}}, \bibinfo {author} {\bibfnamefont
  {T.}~\bibnamefont {Ki{\o}rboe}}, \bibinfo {author} {\bibfnamefont
  {E.}~\bibnamefont {Lauga}}, \bibinfo {author} {\bibfnamefont {W.~C.~K.}\
  \bibnamefont {Poon}}, \bibinfo {author} {\bibfnamefont {A.}~\bibnamefont
  {DeSimone}}, \bibinfo {author} {\bibfnamefont {S.}~\bibnamefont {Mui{\~
  n}os-Landin}}, \bibinfo {author} {\bibfnamefont {A.}~\bibnamefont {Fischer}},
  \bibinfo {author} {\bibfnamefont {N.~A.}\ \bibnamefont {S{\" o}ker}},
  \bibinfo {author} {\bibfnamefont {F.}~\bibnamefont {Cichos}}, \bibinfo
  {author} {\bibfnamefont {R.}~\bibnamefont {Kapral}}, \bibinfo {author}
  {\bibfnamefont {P.}~\bibnamefont {Gaspard}}, \bibinfo {author} {\bibfnamefont
  {M.}~\bibnamefont {Ripoll}}, \bibinfo {author} {\bibfnamefont
  {F.}~\bibnamefont {Sagues}}, \bibinfo {author} {\bibfnamefont
  {A.}~\bibnamefont {Doostmohammadi}}, \bibinfo {author} {\bibfnamefont
  {Y.~M.}\ \bibnamefont {Yeomans}}, \bibinfo {author} {\bibfnamefont {I.~S.}\
  \bibnamefont {Aranson}}, \bibinfo {author} {\bibfnamefont {C.}~\bibnamefont
  {Bechinger}}, \bibinfo {author} {\bibfnamefont {H.}~\bibnamefont {Stark}},
  \bibinfo {author} {\bibfnamefont {C.~K.}\ \bibnamefont {Hemelrijk}}, \bibinfo
  {author} {\bibfnamefont {F.~J.}\ \bibnamefont {Nedelec}}, \bibinfo {author}
  {\bibfnamefont {T.}~\bibnamefont {Sarkar}}, \bibinfo {author} {\bibfnamefont
  {T.}~\bibnamefont {Aryaksama}}, \bibinfo {author} {\bibfnamefont
  {M.}~\bibnamefont {Lacroix}}, \bibinfo {author} {\bibfnamefont
  {G.}~\bibnamefont {Duclos}}, \bibinfo {author} {\bibfnamefont
  {V.}~\bibnamefont {Yashunsky}}, \bibinfo {author} {\bibfnamefont
  {P.}~\bibnamefont {Silberzan}}, \bibinfo {author} {\bibfnamefont
  {M.}~\bibnamefont {Arroyo}}, \ and\ \bibinfo {author} {\bibfnamefont
  {S.}~\bibnamefont {Kale}},\ }\bibfield  {title} {\enquote {\bibinfo {title}
  {The 2020 motile active matter roadmap},}\ }\href {\doibase
  10.1088/1361-648X/ab6348} {\bibfield  {journal} {\bibinfo  {journal} {J.
  Phys.: Condens. Matter}\ }\textbf {\bibinfo {volume} {32}},\ \bibinfo {pages}
  {193001} (\bibinfo {year} {2020})}\BibitemShut {NoStop}%
\bibitem [{\citenamefont {Helbing}(2001)}]{Helbing01}%
  \BibitemOpen
  \bibfield  {author} {\bibinfo {author} {\bibfnamefont {D.}~\bibnamefont
  {Helbing}},\ }\bibfield  {title} {\enquote {\bibinfo {title} {Traffic and
  related self-driven many-particle systems},}\ }\href {\doibase
  10.1103/RevModPhys.73.1067} {\bibfield  {journal} {\bibinfo  {journal} {Rev.
  Mod. Phys.}\ }\textbf {\bibinfo {volume} {73}},\ \bibinfo {pages}
  {1067--1141} (\bibinfo {year} {2001})}\BibitemShut {NoStop}%
\bibitem [{\citenamefont {Wang}\ \emph {et~al.}(2021)\citenamefont {Wang},
  \citenamefont {Phan}, \citenamefont {Li}, \citenamefont {Wombacher},
  \citenamefont {Qu}, \citenamefont {Peng}, \citenamefont {Chen}, \citenamefont
  {Goldman}, \citenamefont {Levin}, \citenamefont {Austin},\ and\ \citenamefont
  {Liu}}]{Wang21}%
  \BibitemOpen
  \bibfield  {author} {\bibinfo {author} {\bibfnamefont {G.}~\bibnamefont
  {Wang}}, \bibinfo {author} {\bibfnamefont {T.~V.}\ \bibnamefont {Phan}},
  \bibinfo {author} {\bibfnamefont {S.}~\bibnamefont {Li}}, \bibinfo {author}
  {\bibfnamefont {M.}~\bibnamefont {Wombacher}}, \bibinfo {author}
  {\bibfnamefont {J.}~\bibnamefont {Qu}}, \bibinfo {author} {\bibfnamefont
  {Y.}~\bibnamefont {Peng}}, \bibinfo {author} {\bibfnamefont {G.}~\bibnamefont
  {Chen}}, \bibinfo {author} {\bibfnamefont {D.~I.}\ \bibnamefont {Goldman}},
  \bibinfo {author} {\bibfnamefont {S.~A.}\ \bibnamefont {Levin}}, \bibinfo
  {author} {\bibfnamefont {R.~H.}\ \bibnamefont {Austin}}, \ and\ \bibinfo
  {author} {\bibfnamefont {L.}~\bibnamefont {Liu}},\ }\bibfield  {title}
  {\enquote {\bibinfo {title} {Emergent field-driven robot swarm states},}\
  }\href {\doibase 10.1103/PhysRevLett.126.108002} {\bibfield  {journal}
  {\bibinfo  {journal} {Phys. Rev. Lett.}\ }\textbf {\bibinfo {volume} {126}},\
  \bibinfo {pages} {108002} (\bibinfo {year} {2021})}\BibitemShut {NoStop}%
\bibitem [{\citenamefont {Ben~Zion}\ \emph {et~al.}(2023)\citenamefont
  {Ben~Zion}, \citenamefont {Fersula}, \citenamefont {Bredeche},\ and\
  \citenamefont {Dauchot}}]{BenZion23}%
  \BibitemOpen
  \bibfield  {author} {\bibinfo {author} {\bibfnamefont {M.~Y.}\ \bibnamefont
  {Ben~Zion}}, \bibinfo {author} {\bibfnamefont {J.}~\bibnamefont {Fersula}},
  \bibinfo {author} {\bibfnamefont {N.}~\bibnamefont {Bredeche}}, \ and\
  \bibinfo {author} {\bibfnamefont {O.}~\bibnamefont {Dauchot}},\ }\bibfield
  {title} {\enquote {\bibinfo {title} {Morphological computation and
  decentralized learning in a swarm of sterically interacting robots},}\ }\href
  {\doibase 10.1126/scirobotics.abo6140} {\bibfield  {journal} {\bibinfo
  {journal} {Sci. Robotics}\ }\textbf {\bibinfo {volume} {8}},\ \bibinfo
  {pages} {eabo6140} (\bibinfo {year} {2023})}\BibitemShut {NoStop}%
\bibitem [{\citenamefont {Solon}\ \emph {et~al.}(2015)\citenamefont {Solon},
  \citenamefont {Fily}, \citenamefont {Baskaran}, \citenamefont {Cates},
  \citenamefont {Kafri}, \citenamefont {Kardar},\ and\ \citenamefont
  {Tailleur}}]{Solon15}%
  \BibitemOpen
  \bibfield  {author} {\bibinfo {author} {\bibfnamefont {A.~P.}\ \bibnamefont
  {Solon}}, \bibinfo {author} {\bibfnamefont {Y.}~\bibnamefont {Fily}},
  \bibinfo {author} {\bibfnamefont {A.}~\bibnamefont {Baskaran}}, \bibinfo
  {author} {\bibfnamefont {M.~E.}\ \bibnamefont {Cates}}, \bibinfo {author}
  {\bibfnamefont {Y.}~\bibnamefont {Kafri}}, \bibinfo {author} {\bibfnamefont
  {M.}~\bibnamefont {Kardar}}, \ and\ \bibinfo {author} {\bibfnamefont
  {J.}~\bibnamefont {Tailleur}},\ }\bibfield  {title} {\enquote {\bibinfo
  {title} {Pressure is not a state function for generic active fluids},}\
  }\href {\doibase 10.1038/NPHYS3377} {\bibfield  {journal} {\bibinfo
  {journal} {Nature Phys.}\ }\textbf {\bibinfo {volume} {11}},\ \bibinfo
  {pages} {673--678} (\bibinfo {year} {2015})}\BibitemShut {NoStop}%
\bibitem [{\citenamefont {Takatori}\ \emph {et~al.}(2014)\citenamefont
  {Takatori}, \citenamefont {Yan},\ and\ \citenamefont {Brady}}]{Takatori14}%
  \BibitemOpen
  \bibfield  {author} {\bibinfo {author} {\bibfnamefont {S.~C.}\ \bibnamefont
  {Takatori}}, \bibinfo {author} {\bibfnamefont {W.}~\bibnamefont {Yan}}, \
  and\ \bibinfo {author} {\bibfnamefont {J.~F.}\ \bibnamefont {Brady}},\
  }\bibfield  {title} {\enquote {\bibinfo {title} {Swim pressure: Stress
  generation in active matter},}\ }\href {\doibase
  10.1103/PhysRevLett.113.028103} {\bibfield  {journal} {\bibinfo  {journal}
  {Phys. Rev. Lett.}\ }\textbf {\bibinfo {volume} {113}},\ \bibinfo {pages}
  {028103} (\bibinfo {year} {2014})}\BibitemShut {NoStop}%
\bibitem [{\citenamefont {Dabelow}\ \emph {et~al.}(2019)\citenamefont
  {Dabelow}, \citenamefont {Bo},\ and\ \citenamefont {Eichhorn}}]{Dabelow19}%
  \BibitemOpen
  \bibfield  {author} {\bibinfo {author} {\bibfnamefont {L.}~\bibnamefont
  {Dabelow}}, \bibinfo {author} {\bibfnamefont {S.}~\bibnamefont {Bo}}, \ and\
  \bibinfo {author} {\bibfnamefont {R.}~\bibnamefont {Eichhorn}},\ }\bibfield
  {title} {\enquote {\bibinfo {title} {Irreversibility in active matter
  systems: Fluctuation theorem and mutual information},}\ }\href {\doibase
  10.1103/PhysRevX.9.021009} {\bibfield  {journal} {\bibinfo  {journal} {Phys.
  Rev. X}\ }\textbf {\bibinfo {volume} {9}},\ \bibinfo {pages} {021009}
  (\bibinfo {year} {2019})}\BibitemShut {NoStop}%
\bibitem [{\citenamefont {O'Byrne}\ \emph {et~al.}(2022)\citenamefont
  {O'Byrne}, \citenamefont {Kafri}, \citenamefont {Tailleur},\ and\
  \citenamefont {van Wijland}}]{OByrne22}%
  \BibitemOpen
  \bibfield  {author} {\bibinfo {author} {\bibfnamefont {J.}~\bibnamefont
  {O'Byrne}}, \bibinfo {author} {\bibfnamefont {Y.}~\bibnamefont {Kafri}},
  \bibinfo {author} {\bibfnamefont {J.}~\bibnamefont {Tailleur}}, \ and\
  \bibinfo {author} {\bibfnamefont {F.}~\bibnamefont {van Wijland}},\
  }\bibfield  {title} {\enquote {\bibinfo {title} {Time irreversibility in
  active matter, from micro to macro},}\ }\href {\doibase
  10.1038/s42254-021-00406-2} {\bibfield  {journal} {\bibinfo  {journal}
  {Nature Rev. Phys.}\ }\textbf {\bibinfo {volume} {4}},\ \bibinfo {pages}
  {167--183} (\bibinfo {year} {2022})}\BibitemShut {NoStop}%
\bibitem [{\citenamefont {Ray}\ \emph {et~al.}(2014)\citenamefont {Ray},
  \citenamefont {Reichhardt},\ and\ \citenamefont {Reichhardt}}]{Ray14}%
  \BibitemOpen
  \bibfield  {author} {\bibinfo {author} {\bibfnamefont {D.}~\bibnamefont
  {Ray}}, \bibinfo {author} {\bibfnamefont {C.}~\bibnamefont {Reichhardt}}, \
  and\ \bibinfo {author} {\bibfnamefont {C.~J.~Olson}\ \bibnamefont
  {Reichhardt}},\ }\bibfield  {title} {\enquote {\bibinfo {title} {Casimir
  effect in active matter systems},}\ }\href {\doibase
  10.1103/PhysRevE.90.013019} {\bibfield  {journal} {\bibinfo  {journal} {Phys.
  Rev. E}\ }\textbf {\bibinfo {volume} {90}},\ \bibinfo {pages} {013019}
  (\bibinfo {year} {2014})}\BibitemShut {NoStop}%
\bibitem [{\citenamefont {Ni}\ \emph {et~al.}(2015)\citenamefont {Ni},
  \citenamefont {Cohen~Stuart},\ and\ \citenamefont {Bolhuis}}]{Ni15}%
  \BibitemOpen
  \bibfield  {author} {\bibinfo {author} {\bibfnamefont {R.}~\bibnamefont
  {Ni}}, \bibinfo {author} {\bibfnamefont {M.~A.}\ \bibnamefont
  {Cohen~Stuart}}, \ and\ \bibinfo {author} {\bibfnamefont {P.~G.}\
  \bibnamefont {Bolhuis}},\ }\bibfield  {title} {\enquote {\bibinfo {title}
  {Tunable long range forces mediated by self-propelled colloidal hard
  spheres},}\ }\href {\doibase 10.1103/PhysRevLett.114.018302} {\bibfield
  {journal} {\bibinfo  {journal} {Phys. Rev. Lett.}\ }\textbf {\bibinfo
  {volume} {114}},\ \bibinfo {pages} {018302} (\bibinfo {year}
  {2015})}\BibitemShut {NoStop}%
\bibitem [{\citenamefont {Yan}\ and\ \citenamefont {Brady}(2015)}]{Yan15}%
  \BibitemOpen
  \bibfield  {author} {\bibinfo {author} {\bibfnamefont {W.}~\bibnamefont
  {Yan}}\ and\ \bibinfo {author} {\bibfnamefont {J.~F.}\ \bibnamefont
  {Brady}},\ }\bibfield  {title} {\enquote {\bibinfo {title} {The force on a
  boundary in active matter},}\ }\href {\doibase 10.1017/jfm.2015.621}
  {\bibfield  {journal} {\bibinfo  {journal} {J. Fluid Mech.}\ }\textbf
  {\bibinfo {volume} {785}},\ \bibinfo {pages} {R1} (\bibinfo {year}
  {2015})}\BibitemShut {NoStop}%
\bibitem [{\citenamefont {Wagner}\ \emph {et~al.}(2022)\citenamefont {Wagner},
  \citenamefont {Hagan},\ and\ \citenamefont {Baskaran}}]{Wagner22}%
  \BibitemOpen
  \bibfield  {author} {\bibinfo {author} {\bibfnamefont {C.~G.}\ \bibnamefont
  {Wagner}}, \bibinfo {author} {\bibfnamefont {M.~F.}\ \bibnamefont {Hagan}}, \
  and\ \bibinfo {author} {\bibfnamefont {A.}~\bibnamefont {Baskaran}},\
  }\bibfield  {title} {\enquote {\bibinfo {title} {Steady states of active
  {B}rownian particles interacting with boundaries},}\ }\href {\doibase
  10.1088/1742-5468/ac42cf} {\bibfield  {journal} {\bibinfo  {journal} {J.
  Stat. Mech.}\ }\textbf {\bibinfo {volume} {2022}},\ \bibinfo {pages} {013208}
  (\bibinfo {year} {2022})}\BibitemShut {NoStop}%
\bibitem [{\citenamefont {Speck}(2020)}]{Speck20}%
  \BibitemOpen
  \bibfield  {author} {\bibinfo {author} {\bibfnamefont {T.}~\bibnamefont
  {Speck}},\ }\bibfield  {title} {\enquote {\bibinfo {title} {Collective forces
  in scalar active matter},}\ }\href {\doibase 10.1039/DoSM00176G} {\bibfield
  {journal} {\bibinfo  {journal} {Soft Matter}\ }\textbf {\bibinfo {volume}
  {16}},\ \bibinfo {pages} {2652} (\bibinfo {year} {2020})}\BibitemShut
  {NoStop}%
\bibitem [{\citenamefont {Reichhardt}\ and\ \citenamefont
  {Reichhardt}(2017)}]{Reichhardt17a}%
  \BibitemOpen
  \bibfield  {author} {\bibinfo {author} {\bibfnamefont {C.~J.~Olson}\
  \bibnamefont {Reichhardt}}\ and\ \bibinfo {author} {\bibfnamefont
  {C.}~\bibnamefont {Reichhardt}},\ }\bibfield  {title} {\enquote {\bibinfo
  {title} {Ratchet effects in active matter systems},}\ }\href {\doibase
  10.1146/annurev-conmatphys-031016-025522} {\bibfield  {journal} {\bibinfo
  {journal} {Ann. Rev. Condens. Matter Phys.}\ }\textbf {\bibinfo {volume}
  {8}},\ \bibinfo {pages} {51--75} (\bibinfo {year} {2017})}\BibitemShut
  {NoStop}%
\bibitem [{\citenamefont {Fily}\ and\ \citenamefont
  {Marchetti}(2012)}]{Fily12}%
  \BibitemOpen
  \bibfield  {author} {\bibinfo {author} {\bibfnamefont {Y.}~\bibnamefont
  {Fily}}\ and\ \bibinfo {author} {\bibfnamefont {M.~C.}\ \bibnamefont
  {Marchetti}},\ }\bibfield  {title} {\enquote {\bibinfo {title} {Athermal
  phase separation of self-propelled particles with no alignment},}\ }\href
  {\doibase 10.1103/PhysRevLett.108.235702} {\bibfield  {journal} {\bibinfo
  {journal} {Phys. Rev. Lett.}\ }\textbf {\bibinfo {volume} {108}},\ \bibinfo
  {pages} {235702} (\bibinfo {year} {2012})}\BibitemShut {NoStop}%
\bibitem [{\citenamefont {Redner}\ \emph {et~al.}(2013)\citenamefont {Redner},
  \citenamefont {Hagan},\ and\ \citenamefont {Baskaran}}]{Redner13}%
  \BibitemOpen
  \bibfield  {author} {\bibinfo {author} {\bibfnamefont {G.~S.}\ \bibnamefont
  {Redner}}, \bibinfo {author} {\bibfnamefont {M.~F.}\ \bibnamefont {Hagan}}, \
  and\ \bibinfo {author} {\bibfnamefont {A.}~\bibnamefont {Baskaran}},\
  }\bibfield  {title} {\enquote {\bibinfo {title} {Structure and dynamics of a
  phase-separating active colloidal fluid},}\ }\href {\doibase
  10.1103/PhysRevLett.110.055701} {\bibfield  {journal} {\bibinfo  {journal}
  {Phys. Rev. Lett.}\ }\textbf {\bibinfo {volume} {110}},\ \bibinfo {pages}
  {055701} (\bibinfo {year} {2013})}\BibitemShut {NoStop}%
\bibitem [{\citenamefont {Palacci}\ \emph {et~al.}(2013)\citenamefont
  {Palacci}, \citenamefont {Sacanna}, \citenamefont {Steinberg}, \citenamefont
  {Pine},\ and\ \citenamefont {Chaikin}}]{Palacci13}%
  \BibitemOpen
  \bibfield  {author} {\bibinfo {author} {\bibfnamefont {J.}~\bibnamefont
  {Palacci}}, \bibinfo {author} {\bibfnamefont {S.}~\bibnamefont {Sacanna}},
  \bibinfo {author} {\bibfnamefont {A.~P.}\ \bibnamefont {Steinberg}}, \bibinfo
  {author} {\bibfnamefont {D.~J.}\ \bibnamefont {Pine}}, \ and\ \bibinfo
  {author} {\bibfnamefont {P.~M.}\ \bibnamefont {Chaikin}},\ }\bibfield
  {title} {\enquote {\bibinfo {title} {Living crystals of light-activated
  colloidal surfers},}\ }\href {\doibase 10.1126/science.1230020} {\bibfield
  {journal} {\bibinfo  {journal} {Science}\ }\textbf {\bibinfo {volume}
  {339}},\ \bibinfo {pages} {936--940} (\bibinfo {year} {2013})}\BibitemShut
  {NoStop}%
\bibitem [{\citenamefont {Buttinoni}\ \emph {et~al.}(2013)\citenamefont
  {Buttinoni}, \citenamefont {Bialk\'e}, \citenamefont {K\"ummel},
  \citenamefont {L\"owen}, \citenamefont {Bechinger},\ and\ \citenamefont
  {Speck}}]{Buttinoni13}%
  \BibitemOpen
  \bibfield  {author} {\bibinfo {author} {\bibfnamefont {I.}~\bibnamefont
  {Buttinoni}}, \bibinfo {author} {\bibfnamefont {J.}~\bibnamefont {Bialk\'e}},
  \bibinfo {author} {\bibfnamefont {F.}~\bibnamefont {K\"ummel}}, \bibinfo
  {author} {\bibfnamefont {H.}~\bibnamefont {L\"owen}}, \bibinfo {author}
  {\bibfnamefont {C.}~\bibnamefont {Bechinger}}, \ and\ \bibinfo {author}
  {\bibfnamefont {T.}~\bibnamefont {Speck}},\ }\bibfield  {title} {\enquote
  {\bibinfo {title} {Dynamical clustering and phase separation in suspensions
  of self-propelled colloidal particles},}\ }\href {\doibase
  10.1103/PhysRevLett.110.238301} {\bibfield  {journal} {\bibinfo  {journal}
  {Phys. Rev. Lett.}\ }\textbf {\bibinfo {volume} {110}},\ \bibinfo {pages}
  {238301} (\bibinfo {year} {2013})}\BibitemShut {NoStop}%
\bibitem [{\citenamefont {Cates}\ and\ \citenamefont
  {Tailleur}(2015)}]{Cates15}%
  \BibitemOpen
  \bibfield  {author} {\bibinfo {author} {\bibfnamefont {M.~E.}\ \bibnamefont
  {Cates}}\ and\ \bibinfo {author} {\bibfnamefont {J.}~\bibnamefont
  {Tailleur}},\ }\bibfield  {title} {\enquote {\bibinfo {title}
  {Motility-induced phase separation},}\ }\href {\doibase
  10.1146/annurev-conmatphys-031214-014710} {\bibfield  {journal} {\bibinfo
  {journal} {Annual Review of Condensed Matter Physics}\ }\textbf {\bibinfo
  {volume} {6}},\ \bibinfo {pages} {219--244} (\bibinfo {year}
  {2015})}\BibitemShut {NoStop}%
\bibitem [{\citenamefont {Paoluzzi}\ \emph {et~al.}(2022)\citenamefont
  {Paoluzzi}, \citenamefont {Levis},\ and\ \citenamefont
  {Pagonabarraga}}]{Paoluzzi22}%
  \BibitemOpen
  \bibfield  {author} {\bibinfo {author} {\bibfnamefont {M.}~\bibnamefont
  {Paoluzzi}}, \bibinfo {author} {\bibfnamefont {D.}~\bibnamefont {Levis}}, \
  and\ \bibinfo {author} {\bibfnamefont {I.}~\bibnamefont {Pagonabarraga}},\
  }\bibfield  {title} {\enquote {\bibinfo {title} {From motility-induced
  phase-separation to glassiness in dense active matter},}\ }\href {\doibase
  10.1038/s42005-022-00886-3} {\bibfield  {journal} {\bibinfo  {journal}
  {Commun. Phys.}\ }\textbf {\bibinfo {volume} {5}},\ \bibinfo {pages} {111}
  (\bibinfo {year} {2022})}\BibitemShut {NoStop}%
\bibitem [{\citenamefont {Ni}\ \emph {et~al.}(2014)\citenamefont {Ni},
  \citenamefont {Cohen~Stuart}, \citenamefont {Dijkstra},\ and\ \citenamefont
  {Bolhuis}}]{Ni14}%
  \BibitemOpen
  \bibfield  {author} {\bibinfo {author} {\bibfnamefont {R.}~\bibnamefont
  {Ni}}, \bibinfo {author} {\bibfnamefont {M.~A.}\ \bibnamefont
  {Cohen~Stuart}}, \bibinfo {author} {\bibfnamefont {M.}~\bibnamefont
  {Dijkstra}}, \ and\ \bibinfo {author} {\bibfnamefont {P.~G.}\ \bibnamefont
  {Bolhuis}},\ }\bibfield  {title} {\enquote {\bibinfo {title} {Crystallizing
  hard-sphere glasses by doping with active particles},}\ }\href {\doibase
  10.1039/C4SM01015A} {\bibfield  {journal} {\bibinfo  {journal} {Soft Matter}\
  }\textbf {\bibinfo {volume} {10}},\ \bibinfo {pages} {6609} (\bibinfo {year}
  {2014})}\BibitemShut {NoStop}%
\bibitem [{\citenamefont {K{\" u}mmel}\ \emph {et~al.}(2015)\citenamefont {K{\"
  u}mmel}, \citenamefont {Shabestari}, \citenamefont {Lozano}, \citenamefont
  {Volpe},\ and\ \citenamefont {Bechinger}}]{Kummel15}%
  \BibitemOpen
  \bibfield  {author} {\bibinfo {author} {\bibfnamefont {F.}~\bibnamefont {K{\"
  u}mmel}}, \bibinfo {author} {\bibfnamefont {P.}~\bibnamefont {Shabestari}},
  \bibinfo {author} {\bibfnamefont {C.}~\bibnamefont {Lozano}}, \bibinfo
  {author} {\bibfnamefont {G.}~\bibnamefont {Volpe}}, \ and\ \bibinfo {author}
  {\bibfnamefont {C.}~\bibnamefont {Bechinger}},\ }\bibfield  {title} {\enquote
  {\bibinfo {title} {Formation, compression and surface melting of colloidal
  clusters by active particles},}\ }\href {\doibase 10.1039/c5sm00827a}
  {\bibfield  {journal} {\bibinfo  {journal} {Soft Matter}\ }\textbf {\bibinfo
  {volume} {11}},\ \bibinfo {pages} {6187--6191} (\bibinfo {year}
  {2015})}\BibitemShut {NoStop}%
\bibitem [{\citenamefont {Massana-Cid}\ \emph {et~al.}(2018)\citenamefont
  {Massana-Cid}, \citenamefont {Codina}, \citenamefont {Pagonabarraga},\ and\
  \citenamefont {Tierno}}]{MassanaCid18}%
  \BibitemOpen
  \bibfield  {author} {\bibinfo {author} {\bibfnamefont {H.}~\bibnamefont
  {Massana-Cid}}, \bibinfo {author} {\bibfnamefont {J.}~\bibnamefont {Codina}},
  \bibinfo {author} {\bibfnamefont {I.}~\bibnamefont {Pagonabarraga}}, \ and\
  \bibinfo {author} {\bibfnamefont {P.}~\bibnamefont {Tierno}},\ }\bibfield
  {title} {\enquote {\bibinfo {title} {Active apolar doping determines routes
  to colloidal clusters and gels},}\ }\href {\doibase 10.1073/pnas.1811225115}
  {\bibfield  {journal} {\bibinfo  {journal} {Proc. Natl. Acad. Sci. (USA)}\
  }\textbf {\bibinfo {volume} {115}},\ \bibinfo {pages} {10618--10623}
  (\bibinfo {year} {2018})}\BibitemShut {NoStop}%
\bibitem [{\citenamefont {Omar}\ \emph {et~al.}(2019)\citenamefont {Omar},
  \citenamefont {Wu}, \citenamefont {Wang},\ and\ \citenamefont
  {Brady}}]{Omar19}%
  \BibitemOpen
  \bibfield  {author} {\bibinfo {author} {\bibfnamefont {A.~K.}\ \bibnamefont
  {Omar}}, \bibinfo {author} {\bibfnamefont {Y.}~\bibnamefont {Wu}}, \bibinfo
  {author} {\bibfnamefont {Z.~G.}\ \bibnamefont {Wang}}, \ and\ \bibinfo
  {author} {\bibfnamefont {J.~F.}\ \bibnamefont {Brady}},\ }\bibfield  {title}
  {\enquote {\bibinfo {title} {Swimming to stability: Structural and dynamical
  control via active doping},}\ }\href {\doibase 10.1021/acsnano.8b07421}
  {\bibfield  {journal} {\bibinfo  {journal} {ACS Nano}\ }\textbf {\bibinfo
  {volume} {13}},\ \bibinfo {pages} {560} (\bibinfo {year} {2019})}\BibitemShut
  {NoStop}%
\bibitem [{\citenamefont {Ramananarivo}\ \emph {et~al.}(2019)\citenamefont
  {Ramananarivo}, \citenamefont {Ducrot},\ and\ \citenamefont
  {Palacci}}]{Ramananarivo19}%
  \BibitemOpen
  \bibfield  {author} {\bibinfo {author} {\bibfnamefont {S.}~\bibnamefont
  {Ramananarivo}}, \bibinfo {author} {\bibfnamefont {E.}~\bibnamefont
  {Ducrot}}, \ and\ \bibinfo {author} {\bibfnamefont {J.}~\bibnamefont
  {Palacci}},\ }\bibfield  {title} {\enquote {\bibinfo {title}
  {Activity-controlled annealing of colloidal monolayers},}\ }\href {\doibase
  10.1038/s41467-019-11362-y} {\bibfield  {journal} {\bibinfo  {journal}
  {Nature Commun.}\ }\textbf {\bibinfo {volume} {10}},\ \bibinfo {pages} {3380}
  (\bibinfo {year} {2019})}\BibitemShut {NoStop}%
\bibitem [{\citenamefont {Stenhammar}\ \emph {et~al.}(2015)\citenamefont
  {Stenhammar}, \citenamefont {Wittkowski}, \citenamefont {Marenduzzo},\ and\
  \citenamefont {Cates}}]{Stenhammar15}%
  \BibitemOpen
  \bibfield  {author} {\bibinfo {author} {\bibfnamefont {J.}~\bibnamefont
  {Stenhammar}}, \bibinfo {author} {\bibfnamefont {R.}~\bibnamefont
  {Wittkowski}}, \bibinfo {author} {\bibfnamefont {D.}~\bibnamefont
  {Marenduzzo}}, \ and\ \bibinfo {author} {\bibfnamefont {M.~E.}\ \bibnamefont
  {Cates}},\ }\bibfield  {title} {\enquote {\bibinfo {title} {Activity-induced
  phase separation and self-assembly in mixtures of active and passive
  particles},}\ }\href {\doibase 10.1103/PhysRevLett.114.018301} {\bibfield
  {journal} {\bibinfo  {journal} {Phys. Rev. Lett.}\ }\textbf {\bibinfo
  {volume} {114}},\ \bibinfo {pages} {018301} (\bibinfo {year}
  {2015})}\BibitemShut {NoStop}%
\bibitem [{\citenamefont {Gokhale}\ \emph {et~al.}(2022)\citenamefont
  {Gokhale}, \citenamefont {Li}, \citenamefont {Solon}, \citenamefont {Gore},\
  and\ \citenamefont {Fakhri}}]{Gokhale22}%
  \BibitemOpen
  \bibfield  {author} {\bibinfo {author} {\bibfnamefont {S.}~\bibnamefont
  {Gokhale}}, \bibinfo {author} {\bibfnamefont {J.}~\bibnamefont {Li}},
  \bibinfo {author} {\bibfnamefont {A.}~\bibnamefont {Solon}}, \bibinfo
  {author} {\bibfnamefont {J.}~\bibnamefont {Gore}}, \ and\ \bibinfo {author}
  {\bibfnamefont {N.}~\bibnamefont {Fakhri}},\ }\bibfield  {title} {\enquote
  {\bibinfo {title} {Dynamic clustering of passive colloids in dense
  suspensions of motile bacteria},}\ }\href {\doibase
  10.1103/PhysRevE.105.054605} {\bibfield  {journal} {\bibinfo  {journal}
  {Phys. Rev. E}\ }\textbf {\bibinfo {volume} {105}},\ \bibinfo {pages}
  {054605} (\bibinfo {year} {2022})}\BibitemShut {NoStop}%
\bibitem [{\citenamefont {Paoluzzi}\ \emph {et~al.}(2020)\citenamefont
  {Paoluzzi}, \citenamefont {Leoni},\ and\ \citenamefont
  {Marchetti}}]{Paoluzzi20}%
  \BibitemOpen
  \bibfield  {author} {\bibinfo {author} {\bibfnamefont {M.}~\bibnamefont
  {Paoluzzi}}, \bibinfo {author} {\bibfnamefont {M.}~\bibnamefont {Leoni}}, \
  and\ \bibinfo {author} {\bibfnamefont {M.~C.}\ \bibnamefont {Marchetti}},\
  }\bibfield  {title} {\enquote {\bibinfo {title} {Information and motility
  exchange in collectives of active particles},}\ }\href {\doibase
  10.1039/d0sm00204f} {\bibfield  {journal} {\bibinfo  {journal} {Soft Matter}\
  }\textbf {\bibinfo {volume} {16}},\ \bibinfo {pages} {6317} (\bibinfo {year}
  {2020})}\BibitemShut {NoStop}%
\bibitem [{\citenamefont {Zhao}\ \emph {et~al.}(2022)\citenamefont {Zhao},
  \citenamefont {Huepe},\ and\ \citenamefont {Romanczuk}}]{Zhao22}%
  \BibitemOpen
  \bibfield  {author} {\bibinfo {author} {\bibfnamefont {Y.}~\bibnamefont
  {Zhao}}, \bibinfo {author} {\bibfnamefont {C.}~\bibnamefont {Huepe}}, \ and\
  \bibinfo {author} {\bibfnamefont {P.}~\bibnamefont {Romanczuk}},\ }\bibfield
  {title} {\enquote {\bibinfo {title} {Contagion dynamics in self-organized
  systems of self-propelled agents},}\ }\href {\doibase
  10.1038/s41598-022-06083-0} {\bibfield  {journal} {\bibinfo  {journal} {Sci.
  Rep.}\ }\textbf {\bibinfo {volume} {12}},\ \bibinfo {pages} {2588} (\bibinfo
  {year} {2022})}\BibitemShut {NoStop}%
\bibitem [{\citenamefont {Forg{\' a}cs}\ \emph {et~al.}(2022)\citenamefont
  {Forg{\' a}cs}, \citenamefont {Lib{\' a}l}, \citenamefont {Reichhardt},
  \citenamefont {Hengartner},\ and\ \citenamefont {Reichhardt}}]{Forgacs22}%
  \BibitemOpen
  \bibfield  {author} {\bibinfo {author} {\bibfnamefont {P.}~\bibnamefont
  {Forg{\' a}cs}}, \bibinfo {author} {\bibfnamefont {A.}~\bibnamefont {Lib{\'
  a}l}}, \bibinfo {author} {\bibfnamefont {C.}~\bibnamefont {Reichhardt}},
  \bibinfo {author} {\bibfnamefont {N.}~\bibnamefont {Hengartner}}, \ and\
  \bibinfo {author} {\bibfnamefont {C.~J.~O.}\ \bibnamefont {Reichhardt}},\
  }\bibfield  {title} {\enquote {\bibinfo {title} {Using active matter to
  introduce spatial heterogeneity to the susceptible infected recovered model
  of epidemic spreading},}\ }\href {\doibase 10.1038/s41598-022-15223-5}
  {\bibfield  {journal} {\bibinfo  {journal} {Sci. Rep.}\ }\textbf {\bibinfo
  {volume} {12}},\ \bibinfo {pages} {11229} (\bibinfo {year}
  {2022})}\BibitemShut {NoStop}%
\bibitem [{\citenamefont {Lib\'al}\ \emph {et~al.}(2023)\citenamefont
  {Lib\'al}, \citenamefont {Forg\'acs}, \citenamefont {N\'eda}, \citenamefont
  {Reichhardt}, \citenamefont {Hengartner},\ and\ \citenamefont
  {Reichhardt}}]{Libal23}%
  \BibitemOpen
  \bibfield  {author} {\bibinfo {author} {\bibfnamefont {A.}~\bibnamefont
  {Lib\'al}}, \bibinfo {author} {\bibfnamefont {P.}~\bibnamefont {Forg\'acs}},
  \bibinfo {author} {\bibfnamefont {\'A.}\ \bibnamefont {N\'eda}}, \bibinfo
  {author} {\bibfnamefont {C.}~\bibnamefont {Reichhardt}}, \bibinfo {author}
  {\bibfnamefont {N.}~\bibnamefont {Hengartner}}, \ and\ \bibinfo {author}
  {\bibfnamefont {C.~J.~O.}\ \bibnamefont {Reichhardt}},\ }\bibfield  {title}
  {\enquote {\bibinfo {title} {Transition from susceptible-infected to
  susceptible-infected-recovered dynamics in a
  susceptible-cleric-zombie-recovered active matter model},}\ }\href {\doibase
  10.1103/PhysRevE.107.024604} {\bibfield  {journal} {\bibinfo  {journal}
  {Phys. Rev. E}\ }\textbf {\bibinfo {volume} {107}},\ \bibinfo {pages}
  {024604} (\bibinfo {year} {2023})}\BibitemShut {NoStop}%
\bibitem [{\citenamefont {Tailleur}\ and\ \citenamefont
  {Cates}(2009)}]{Tailleur09}%
  \BibitemOpen
  \bibfield  {author} {\bibinfo {author} {\bibfnamefont {J.}~\bibnamefont
  {Tailleur}}\ and\ \bibinfo {author} {\bibfnamefont {M.~E.}\ \bibnamefont
  {Cates}},\ }\bibfield  {title} {\enquote {\bibinfo {title} {Sedimentation,
  trapping, and rectification of dilute bacteria},}\ }\href {\doibase
  10.1209/0295-5075/86/60002} {\bibfield  {journal} {\bibinfo  {journal} {EPL}\
  }\textbf {\bibinfo {volume} {86}},\ \bibinfo {pages} {60002} (\bibinfo {year}
  {2009})}\BibitemShut {NoStop}%
\bibitem [{\citenamefont {Sep\'ulveda}\ and\ \citenamefont
  {Soto}(2017)}]{Sepulveda17}%
  \BibitemOpen
  \bibfield  {author} {\bibinfo {author} {\bibfnamefont {N.}~\bibnamefont
  {Sep\'ulveda}}\ and\ \bibinfo {author} {\bibfnamefont {R.}~\bibnamefont
  {Soto}},\ }\bibfield  {title} {\enquote {\bibinfo {title} {Wetting
  transitions displayed by persistent active particles},}\ }\href {\doibase
  10.1103/PhysRevLett.119.078001} {\bibfield  {journal} {\bibinfo  {journal}
  {Phys. Rev. Lett.}\ }\textbf {\bibinfo {volume} {119}},\ \bibinfo {pages}
  {078001} (\bibinfo {year} {2017})}\BibitemShut {NoStop}%
\bibitem [{\citenamefont {Neta}\ \emph {et~al.}(2021)\citenamefont {Neta},
  \citenamefont {Tasinkevych}, \citenamefont {Telo~da Gama},\ and\
  \citenamefont {Dias}}]{Neta21}%
  \BibitemOpen
  \bibfield  {author} {\bibinfo {author} {\bibfnamefont {P.~D.}\ \bibnamefont
  {Neta}}, \bibinfo {author} {\bibfnamefont {M.}~\bibnamefont {Tasinkevych}},
  \bibinfo {author} {\bibfnamefont {M.~M.}\ \bibnamefont {Telo~da Gama}}, \
  and\ \bibinfo {author} {\bibfnamefont {C.~S.}\ \bibnamefont {Dias}},\
  }\bibfield  {title} {\enquote {\bibinfo {title} {Wetting of a solid surface
  by active matter},}\ }\href {\doibase 10.1039/D0SM02008G} {\bibfield
  {journal} {\bibinfo  {journal} {Soft Matter}\ }\textbf {\bibinfo {volume}
  {17}},\ \bibinfo {pages} {2468--2478} (\bibinfo {year} {2021})}\BibitemShut
  {NoStop}%
\bibitem [{\citenamefont {Turci}\ and\ \citenamefont
  {Wilding}(2021)}]{Turci21}%
  \BibitemOpen
  \bibfield  {author} {\bibinfo {author} {\bibfnamefont {F.}~\bibnamefont
  {Turci}}\ and\ \bibinfo {author} {\bibfnamefont {N.~B.}\ \bibnamefont
  {Wilding}},\ }\bibfield  {title} {\enquote {\bibinfo {title} {Wetting
  transition of active {B}rownian particles on a thin membrane},}\ }\href
  {\doibase 10.1103/PhysRevLett.127.238002} {\bibfield  {journal} {\bibinfo
  {journal} {Phys. Rev. Lett.}\ }\textbf {\bibinfo {volume} {127}},\ \bibinfo
  {pages} {238002} (\bibinfo {year} {2021})}\BibitemShut {NoStop}%
\bibitem [{\citenamefont {Styles}\ \emph {et~al.}(2021)\citenamefont {Styles},
  \citenamefont {Brown},\ and\ \citenamefont {Sagona}}]{Styles21}%
  \BibitemOpen
  \bibfield  {author} {\bibinfo {author} {\bibfnamefont {K.~M.}\ \bibnamefont
  {Styles}}, \bibinfo {author} {\bibfnamefont {A.~T.}\ \bibnamefont {Brown}}, \
  and\ \bibinfo {author} {\bibfnamefont {A.~P.}\ \bibnamefont {Sagona}},\
  }\bibfield  {title} {\enquote {\bibinfo {title} {A review of using
  mathematical modeling to improve our understanding of bacteriophage,
  bacteria, and eukaryotic interactions},}\ }\href {\doibase
  10.3389/fmicb.2021.724767} {\bibfield  {journal} {\bibinfo  {journal} {Front.
  Microbiol.}\ }\textbf {\bibinfo {volume} {12}},\ \bibinfo {pages} {724767}
  (\bibinfo {year} {2021})}\BibitemShut {NoStop}%
\bibitem [{\citenamefont {Pince}\ \emph {et~al.}(2016)\citenamefont {Pince},
  \citenamefont {Velu}, \citenamefont {Callegari}, \citenamefont {Elahi},
  \citenamefont {Gigan}, \citenamefont {Volpe},\ and\ \citenamefont
  {Volpe}}]{Pince16}%
  \BibitemOpen
  \bibfield  {author} {\bibinfo {author} {\bibfnamefont {E.}~\bibnamefont
  {Pince}}, \bibinfo {author} {\bibfnamefont {S.~K.~P.}\ \bibnamefont {Velu}},
  \bibinfo {author} {\bibfnamefont {A.}~\bibnamefont {Callegari}}, \bibinfo
  {author} {\bibfnamefont {P.}~\bibnamefont {Elahi}}, \bibinfo {author}
  {\bibfnamefont {S.}~\bibnamefont {Gigan}}, \bibinfo {author} {\bibfnamefont
  {G.}~\bibnamefont {Volpe}}, \ and\ \bibinfo {author} {\bibfnamefont
  {G.}~\bibnamefont {Volpe}},\ }\bibfield  {title} {\enquote {\bibinfo {title}
  {Disorder-mediated crowd control in an active matter system},}\ }\href
  {\doibase 10.1038/ncomms10907} {\bibfield  {journal} {\bibinfo  {journal}
  {Nature Commun.}\ }\textbf {\bibinfo {volume} {7}},\ \bibinfo {pages} {10907}
  (\bibinfo {year} {2016})}\BibitemShut {NoStop}%
\bibitem [{\citenamefont {Lavergne}\ \emph {et~al.}(2019)\citenamefont
  {Lavergne}, \citenamefont {Wendehenne}, \citenamefont {Baeuerle},\ and\
  \citenamefont {Bechinger}}]{Lavergne19}%
  \BibitemOpen
  \bibfield  {author} {\bibinfo {author} {\bibfnamefont {F.~A.}\ \bibnamefont
  {Lavergne}}, \bibinfo {author} {\bibfnamefont {H.}~\bibnamefont
  {Wendehenne}}, \bibinfo {author} {\bibfnamefont {T.}~\bibnamefont
  {Baeuerle}}, \ and\ \bibinfo {author} {\bibfnamefont {C.}~\bibnamefont
  {Bechinger}},\ }\bibfield  {title} {\enquote {\bibinfo {title} {Group
  formation and cohesion of active particles with visual perception-dependent
  motility},}\ }\href {\doibase 10.1126/science.aau5347} {\bibfield  {journal}
  {\bibinfo  {journal} {Science}\ }\textbf {\bibinfo {volume} {364}},\ \bibinfo
  {pages} {70} (\bibinfo {year} {2019})}\BibitemShut {NoStop}%
\bibitem [{\citenamefont {B{\" a}uerle}\ \emph {et~al.}(2020)\citenamefont
  {B{\" a}uerle}, \citenamefont {L{\" o}ffler},\ and\ \citenamefont
  {Bechinger}}]{Bauerle20}%
  \BibitemOpen
  \bibfield  {author} {\bibinfo {author} {\bibfnamefont {T.}~\bibnamefont {B{\"
  a}uerle}}, \bibinfo {author} {\bibfnamefont {R.~C.}\ \bibnamefont {L{\"
  o}ffler}}, \ and\ \bibinfo {author} {\bibfnamefont {C.}~\bibnamefont
  {Bechinger}},\ }\bibfield  {title} {\enquote {\bibinfo {title} {Formation of
  stable and responsive collective states in suspensions of active colloids},}\
  }\href {\doibase 10.1038/s41467-020-16161-4} {\bibfield  {journal} {\bibinfo
  {journal} {Nature Commun.}\ }\textbf {\bibinfo {volume} {11}},\ \bibinfo
  {pages} {2547} (\bibinfo {year} {2020})}\BibitemShut {NoStop}%
\bibitem [{\citenamefont {Kermack}\ and\ \citenamefont
  {McKendrick}(1927)}]{Kermack27}%
  \BibitemOpen
  \bibfield  {author} {\bibinfo {author} {\bibfnamefont {W.~O.}\ \bibnamefont
  {Kermack}}\ and\ \bibinfo {author} {\bibfnamefont {A.~G.}\ \bibnamefont
  {McKendrick}},\ }\bibfield  {title} {\enquote {\bibinfo {title} {A
  contribution to the mathematical theory of epidemics},}\ }\href {\doibase
  10.1098/rspa.1927.0118} {\bibfield  {journal} {\bibinfo  {journal} {Proc.
  Roy. Soc. London A}\ }\textbf {\bibinfo {volume} {115}},\ \bibinfo {pages}
  {700} (\bibinfo {year} {1927})}\BibitemShut {NoStop}%
\bibitem [{\citenamefont {Hethcote}(2000)}]{Hethcote00}%
  \BibitemOpen
  \bibfield  {author} {\bibinfo {author} {\bibfnamefont {H.~W.}\ \bibnamefont
  {Hethcote}},\ }\bibfield  {title} {\enquote {\bibinfo {title} {The
  mathematics of infectious diseases},}\ }\href {\doibase
  10.1137/S0036144500371907} {\bibfield  {journal} {\bibinfo  {journal} {SIAM
  Rev.}\ }\textbf {\bibinfo {volume} {42}},\ \bibinfo {pages} {599} (\bibinfo
  {year} {2000})}\BibitemShut {NoStop}%
\bibitem [{\citenamefont {Reichhardt}\ and\ \citenamefont
  {Olson~Reichhardt}(2014)}]{Reichhardt14}%
  \BibitemOpen
  \bibfield  {author} {\bibinfo {author} {\bibfnamefont {C.}~\bibnamefont
  {Reichhardt}}\ and\ \bibinfo {author} {\bibfnamefont {C.~J.}\ \bibnamefont
  {Olson~Reichhardt}},\ }\bibfield  {title} {\enquote {\bibinfo {title} {Active
  matter transport and jamming on disordered landscapes},}\ }\href {\doibase
  10.1103/PhysRevE.90.012701} {\bibfield  {journal} {\bibinfo  {journal} {Phys.
  Rev. E}\ }\textbf {\bibinfo {volume} {90}},\ \bibinfo {pages} {012701}
  (\bibinfo {year} {2014})}\BibitemShut {NoStop}%
\bibitem [{\citenamefont {S\'andor}\ \emph {et~al.}(2017)\citenamefont
  {S\'andor}, \citenamefont {Lib\'al}, \citenamefont {Reichhardt},\ and\
  \citenamefont {Olson~Reichhardt}}]{Sandor17a}%
  \BibitemOpen
  \bibfield  {author} {\bibinfo {author} {\bibfnamefont {Cs.}\ \bibnamefont
  {S\'andor}}, \bibinfo {author} {\bibfnamefont {A.}~\bibnamefont {Lib\'al}},
  \bibinfo {author} {\bibfnamefont {C.}~\bibnamefont {Reichhardt}}, \ and\
  \bibinfo {author} {\bibfnamefont {C.~J.}\ \bibnamefont {Olson~Reichhardt}},\
  }\bibfield  {title} {\enquote {\bibinfo {title} {Dynamic phases of active
  matter systems with quenched disorder},}\ }\href {\doibase
  10.1103/PhysRevE.95.032606} {\bibfield  {journal} {\bibinfo  {journal} {Phys.
  Rev. E}\ }\textbf {\bibinfo {volume} {95}},\ \bibinfo {pages} {032606}
  (\bibinfo {year} {2017})}\BibitemShut {NoStop}%
\bibitem [{\citenamefont {videos are available~at xxx.}()}]{Suppl}%
  \BibitemOpen
  \bibfield  {author} {\bibinfo {author} {\bibfnamefont {Supplementary}\
  \bibnamefont {videos are available~at xxx.}},\ }\href@noop {} {}\BibitemShut
  {NoStop}%
\bibitem [{\citenamefont {Morin}\ \emph {et~al.}(2017)\citenamefont {Morin},
  \citenamefont {Desreumaux}, \citenamefont {Caussin},\ and\ \citenamefont
  {Bartolo}}]{Morin17}%
  \BibitemOpen
  \bibfield  {author} {\bibinfo {author} {\bibfnamefont {A.}~\bibnamefont
  {Morin}}, \bibinfo {author} {\bibfnamefont {N.}~\bibnamefont {Desreumaux}},
  \bibinfo {author} {\bibfnamefont {J.-B.}\ \bibnamefont {Caussin}}, \ and\
  \bibinfo {author} {\bibfnamefont {D.}~\bibnamefont {Bartolo}},\ }\bibfield
  {title} {\enquote {\bibinfo {title} {Distortion and destruction of colloidal
  flocks in disordered environments},}\ }\href {\doibase 10.1038/nphys3903}
  {\bibfield  {journal} {\bibinfo  {journal} {Nature Phys.}\ }\textbf {\bibinfo
  {volume} {13}},\ \bibinfo {pages} {63--67} (\bibinfo {year}
  {2017})}\BibitemShut {NoStop}%
\bibitem [{\citenamefont {Bhattacharjee}\ and\ \citenamefont
  {Dutta}(2019)}]{Bhattacharjee19a}%
  \BibitemOpen
  \bibfield  {author} {\bibinfo {author} {\bibfnamefont {T.}~\bibnamefont
  {Bhattacharjee}}\ and\ \bibinfo {author} {\bibfnamefont {S.~S.}\ \bibnamefont
  {Dutta}},\ }\bibfield  {title} {\enquote {\bibinfo {title} {Bacterial hopping
  and trapping in porous media},}\ }\href {\doibase 10.1038/s41467-019-10115-1}
  {\bibfield  {journal} {\bibinfo  {journal} {Nature Commun.}\ }\textbf
  {\bibinfo {volume} {10}},\ \bibinfo {pages} {2075} (\bibinfo {year}
  {2019})}\BibitemShut {NoStop}%
\bibitem [{\citenamefont {Forg\'acs}\ \emph {et~al.}(2021)\citenamefont
  {Forg\'acs}, \citenamefont {Lib\'al}, \citenamefont {Reichhardt},\ and\
  \citenamefont {Reichhardt}}]{Forgacs21}%
  \BibitemOpen
  \bibfield  {author} {\bibinfo {author} {\bibfnamefont {P.}~\bibnamefont
  {Forg\'acs}}, \bibinfo {author} {\bibfnamefont {A.}~\bibnamefont {Lib\'al}},
  \bibinfo {author} {\bibfnamefont {C.}~\bibnamefont {Reichhardt}}, \ and\
  \bibinfo {author} {\bibfnamefont {C.~J.~O.}\ \bibnamefont {Reichhardt}},\
  }\bibfield  {title} {\enquote {\bibinfo {title} {Active matter shepherding
  and clustering in inhomogeneous environments},}\ }\href {\doibase
  10.1103/PhysRevE.104.044613} {\bibfield  {journal} {\bibinfo  {journal}
  {Phys. Rev. E}\ }\textbf {\bibinfo {volume} {104}},\ \bibinfo {pages}
  {044613} (\bibinfo {year} {2021})}\BibitemShut {NoStop}%
\end{thebibliography}%

\end{document}